\documentstyle[epsf]{article}
\newcommand{\bigzerou}{%
\smash{\lower1.7ex\hbox{\bg 0}}}
\newtheorem{theorem}{Theorem}
\newtheorem{prop}{Proposition}
\newtheorem{defi}{Definition}
\newtheorem{cor}{Corollary}
\newtheorem{conj}{Conjecture}
\newtheorem{lem}{Lemma}

\newcommand{\ba}{\begin{eqnarray}}
\newcommand{\ea}{\end{eqnarray}}
\newcommand{\no}{\nonumber}

\newcommand{\mapright}[1]{%
\smash{\mathop{%
\hbox to 1.0cm{\rightarrowfill}}\limits^{#1}}}
\newcommand{\mapleft}[1]{%
\smash{\mathop{%
\hbox to 1.3cm{\leftarrowfill}}\limits^{#1}}}

\newcommand{\mapup}[1]{\Big\uparrow
\rlap{$\vcenter{\hbox{$\scriptstyle#1\,\;\;$}}$}}

\begin{document}
\title{
\begin{flushright}
%  \begin{minipage}[b]{5em}
%    \normalsize
%             \\
%  \end{minipage}
\end{flushright}
{\bf On the Quantum Cohomology Rings of General Type Projective
Hypersurfaces and Generalized Mirror Transformation}}
\author{{Masao Jinzenji}\\
\\ 
\it Department of Mathematical Sciences\\
\it University of Tokyo\\
\it  Meguro-ku, Tokyo 153-8914, Japan}
\maketitle
\begin{abstract}
In this paper, we study the structure of the quantum cohomology ring 
of a projective hypersurface with non-positive 1st Chern class. 
We prove a theorem which suggests that the mirror transformation of 
the quantum cohomology of a projective Calabi-Yau hypersurface has 
a close relation 
with the ring of symmetric functions, or with Schur polynomials. 
With this result in mind, we propose a generalized mirror 
transformation on the 
quantum cohomology of a hypersurface with negative first Chern class 
and construct an explicit prediction formula for three point Gromov-Witten 
invariants up to cubic rational curves. 
We also construct a projective space 
resolution of the moduli space of polynomial maps, which is in 
a good correspondence with the terms that appear in the generalized 
mirror transformation.      
\footnote{ e-mail 
address: jin@hep-th.phys.s.u-tokyo.ac.jp}  
\end{abstract}
\section{Introduction}
This paper is a continuation of \cite{cj}. Here, we are mainly 
interested in the quantum cohomology ring of a projective hypersurface 
with non-positive 1st Chern class.

Up to now, many beautiful results on the quantum cohomology rings of 
K\"ahler manifolds with non-negative first Chern classes are known.
In particular, A.B.Givental \cite{giv}, B.Kim \cite{kim} and 
B.Lian-K.Liu-S.T. Yau \cite{yau} revealed a deep relation 
between the quantum cohomology of complete intersections in homogeneous 
spaces with non-negative first Chern class and the hypergeometric series. 
Then, what happens in the quantum cohomology of manifolds with negative 
first Chern class? Before we turn into this topic, 
we look back at the quantum cohomology of Calabi-Yau manifolds. 
Roughly speaking, the mirror calculation of the quantum cohomology of 
Calabi-Yau manifolds consists of two parts. 
The first part is the evaluation of the hypergeometric series obtained from 
the B-model of its mirror manifold and the second part is a translation 
of the hypergeometric data (B-model) into the geometrical one (A-model) 
by a coordinate change of the deformation parameter, i.e., the mirror map.
In \cite{giv}, Givental revealed the fact that in the case of 
projective Fano hypersurface with first Chern class $c\cdot H,\;\;(c\geq 2)$,
we can do without the second part of the mirror calculation.  
In \cite{jin}, \cite{mp} ,\cite{yau}, it was argued that the second 
part has a close relation with a difference between the non-linear 
sigma model and the gauged linear sigma model, or , in other words,
a difference between the moduli space of polynomial maps and  
its toric compactification.
An easy dimensional counting leads us to the conclusion that we have the 
same problem in treating the quantum cohomology of a projective hypersurface
with negative first Chern class. 

In this paper, we give some results arguing that 
there exists a generalization of two processes in the 
mirror calculation in the case of  
the quantum cohomology of the general type projective hypersurface.
Our objects that correspond to the hypergeometric series in the case of 
a hypersurface of nonpositive first Chern class is the virtual structure 
constants obtained by directly applying recursive formulas for 
the structure constants of the quantum cohomology of Fano hypersurfaces 
with first Chern class $c\cdot H, \;\; (c\geq 2)$. These recursive
formulas express the structure constants for degree $k$ hypersurface 
($M_{N}^{k}$) in $CP^{N-1}$ in terms of those of $M_{N+1}^{k}$.
By these formulas, we can reduce the dimension of the hypersurface 
conserving its degree.
 We constructed them up to mapping degree $4$ in \cite{cj} (we later
constructed the recursive formulas for rational
curves of mapping degree $5$).
We observed that the virtual structure constants of the 
Calabi-Yau hypersurface 
$M_{k}^{k}$ reproduce the coefficients of the hypergeometric series used in 
the mirror calculation up to mapping degree $5$, and we expect here that 
these constants produce objects like the hypergeometric series, even in 
the case of the general type hypersurfaces. For the second part of the mirror
calculation, we propose an idea for construction of the 
generalized mirror transformation.
In this paper, we argue that this transformation can be effectively 
consturcted from the one for Calabi-Yau hypersurfaces. In \cite{cj}, 
we constructed an implicit formula that translates the virtual structure 
constants into the true ones in the case of the Calabi-Yau hypersurface, 
and here, we prove a theorem that implies 
a close relation between this mirror transformation and Schur polynomials, 
or partitions of integers (Young diagrams). Then we conjecture that the 
structure revealed in our theorem is still conserved in the case of 
the hypersurface with negative 
first Chern class. Indeed, we construct an explicit form of the 
generalized mirror transformation in this case up to mapping degree
$3$ (We also obtained partial results for mapping degree $4$)
with some technical assumption. 
Our formula can also be regarded as 
a prediction formula for the $3$-point (with one insertion of K\"ahler
form ) Gromov-Witten invariant of the general type projective hypersurfaces.

Our idea of generalization of the 
mirror transformation has a geometrical origin.
We construct a set-theoretic projective space resolution of the moduli space
${\cal M}_{d}^{CP^{N-1}}$, that is the moduli space of polynomial maps from 
$CP^{1}$ to $CP^{N-1}$ of mapping degree $d$. The resolution diagram 
consists of the direct product of $CP^{n}\;$ ($n$ varies) and these 
spaces are labeled by partitions of nonnegative integers, or mapping
degree. We observe that there is one-to-one correspondence between
the spaces in the resolution diagram and the terms in the generalized mirror
transformation. On the other hand, as we mentioned in \cite{jin}, 
dimensional count leads us to the conclusion that
we can 
not see the resolution structure in the case of a Fano hypersurface 
with first Chern class $c\cdot H\;\; (c\geq 2)$. 
In sum, we expect a deep relation between 
this resolution and the ring of symmetric functions \cite{mac}.
We think this point of view should be pursued further.  

This paper is organized as follows.
 
In section 2, we construct a projective space resolution of the moduli space 
${\cal M}_{d}^{CP^{N-1}}$ of polynomial maps from $CP^{1}$ to 
$CP^{N-1}$ of mapping degree $d$ and explain its connection with 
partitions of integers.
 
In section 3, we briefly review the results on the K\"ahler sub-ring of 
the quantum cohomology 
ring of a projective hypersurface with nonnegative first Chern class.
We also introduce here the virtual structure constants.
 
In section 4, we construct recursive formulas that represent 
the structure constants of the quantum cohomology of the hypersuface of 
degree $k$ in $CP^{N-1}$ in terms of the ones of the hypersuface
of the same degree in $CP^{N}$ 
in the case of non-positive first Chern class and 
construct the generalized mirror transformation up to mapping degree 2. 
Next, we state a theorem on the mirror transformation of the Calabi-Yau
hypersurfaces and explain its relation with Schur polynomials and with 
the projective space resolution in section 2.

In section 5, which is the main section of this paper, 
we make a general conjecture on the form of 
the generalized mirror transformation using the results in section 4.
Then we construct an explicit formula 
of the transformation in the case of mapping degree $3$, (we also obtained
partial results for mapping degree $4$) using our conjecture. 

In section 6, we show some explicit examples of the quantum cohomology 
of general type  hypersurface $M_{N}^{k}$ for lower $N$ and $k$.
  
\section{Projective Space Resolution of Moduli Space}
Topic of this section is not logically connected to the derivation of the  
generalized mirror transformation, which is the main topic of this paper, 
but it gives us a geometrical back-ground of the conjecture
that will be proposed in section 5. 

Here, we propose a set-theoretic projective space resolution 
of the moduli space ${\cal M}_{d}^{CP^{N-1}}$  
of polynomial maps of mapping degree $d$ from $CP^{1}$ to $CP^{N-1}$. 
Note that 
this moduli space is non-compact and different from the moduli space 
of stable maps from $CP^{1}$ to $CP^{N-1}$.     
This point of view is already discussed by many authors and 
is called ``Gauged Linear Sigma Model''. 

Let us define that the sequence of maps 
\begin{equation}
A\;\mapright{f}\;B\;\mapright{g}\;C
\label{exact}
\end{equation}
is set-theoretically exact if $g$ is a bijective map between 
$B-Im(f)$ and $Im(g)\subset C$. 
If $B-Im(f)$ is the disjoint union of the sets $B_{j},\;(j=1,2,\cdots,m)$
and if the maps $g_{j}:B_{j}\rightarrow C_{j},\;(j=1,2,\cdots,m)$ are 
bijection between $B_{j}$ and $Im(g_{j})$, we say that the branched 
sequence of maps $f:A\rightarrow B$ and $g_{j}:B_{j}\rightarrow C_{j}$
are set-theoretically exact.
Then our  resolution diagrams consist of the branched sequences of maps 
that are set-theoretically exact.
For some lower degrees, the resolution diagrams are given in figure 1.
\begin{figure}[h]
      \epsfxsize=12cm
     \centerline{\epsfbox{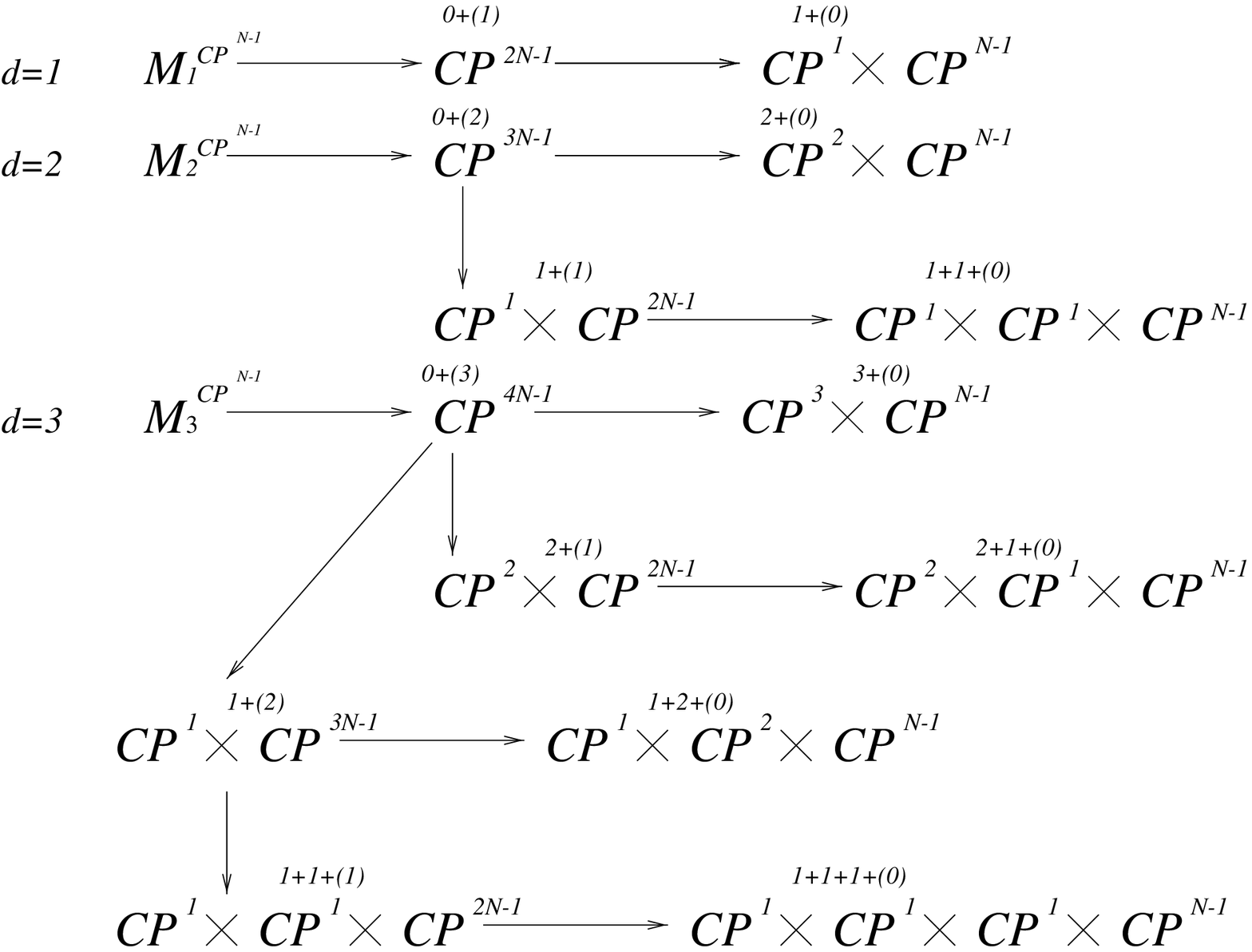}}
    \caption{\bf Projective Space Resolution of ${\cal M}_{d}^{CP^{N-1}}$ 
for $d=1,2,3$}
\end{figure}
Let us explain the inductive construction of these diagrams.
First, take an injective map $\iota$ 
from ${\cal M}_{d}^{CP^{N-1}}$           
to $CP^{(d+1)N-1}$ in figure 2.
 The map $\iota$ is defined as follows. 
\begin{equation}
\iota((\sum_{j=0}^{d}a_{j}^{1}s^{j}t^{d-j}:\cdots:
\sum_{j=0}^{d}a_{j}^{N}s^{j}t^{d-j}))=(a_{0}^{1}:\cdots:a_{d}^{1}:a_{0}^{2}:\cdots:a_{d}^{2}:\cdots:
a_{0}^{N}:\cdots:a_{d}^{N})
\label{toric}
\end{equation}
\begin{figure}[h]
      \epsfxsize=4cm
     \centerline{\epsfbox{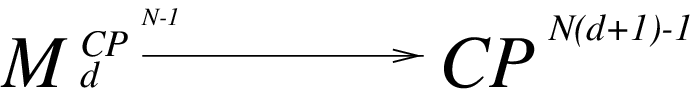}}
     \caption{\bf Injective Map}
\end{figure}
Note that homogeneous polynomials of degree $d$ 
$\sum_{j=0}^{d}a_{j}^{k}s^{j}t^{d-j}\;(k=1,\cdots,N)$ must not have common 
divisor of positive degree, because otherwise they correspond to maps of
lower degree by a
projective equivalence.
Obviously, $CP^{N(d+1)-1}-Im(\iota)$ is bijective to the disjoint union of 
$CP^{j}\times {\cal M}_{d-j}^{CP^{N-1}}\;\; (j=1,2,\cdots, d)$ 
by the following bijection:
\begin{eqnarray}
&&\eta_{j}^{d}((\sum_{k=0}^{j}c_{k}s^{k}t^{j-k})\times(\sum_{k=0}^{d-j}
a_{k}^{1}s^{k}t^{d-j-k}:\cdots:
\sum_{k=0}^{d-j}a_{k}^{N}s^{k}t^{d-j-k}))\no\\ 
&&:=\iota(((\sum_{k=0}^{d-j}
a_{k}^{1}s^{k}t^{d-j-k})\cdot(\sum_{l=0}^{j}c_{l}s^{l}t^{j-l}):\cdots:
(\sum_{k=0}^{d-j}a_{k}^{N}s^{k}t^{d-j-k})
\cdot(\sum_{l=0}^{j}c_{l}s^{l}t^{j-l}))).
\label{bd}
\end{eqnarray}
Here, we extend the definition of $\iota$ to the general polynomial maps 
having a common divisor of positive degree.
Combining $\iota$ and $(\eta_{j}^{d})^{-1}$, we obtain a set-theoretic 
short exact sequence in figure 3.
\begin{figure}[h]
      \epsfxsize=9cm
     \centerline{\epsfbox{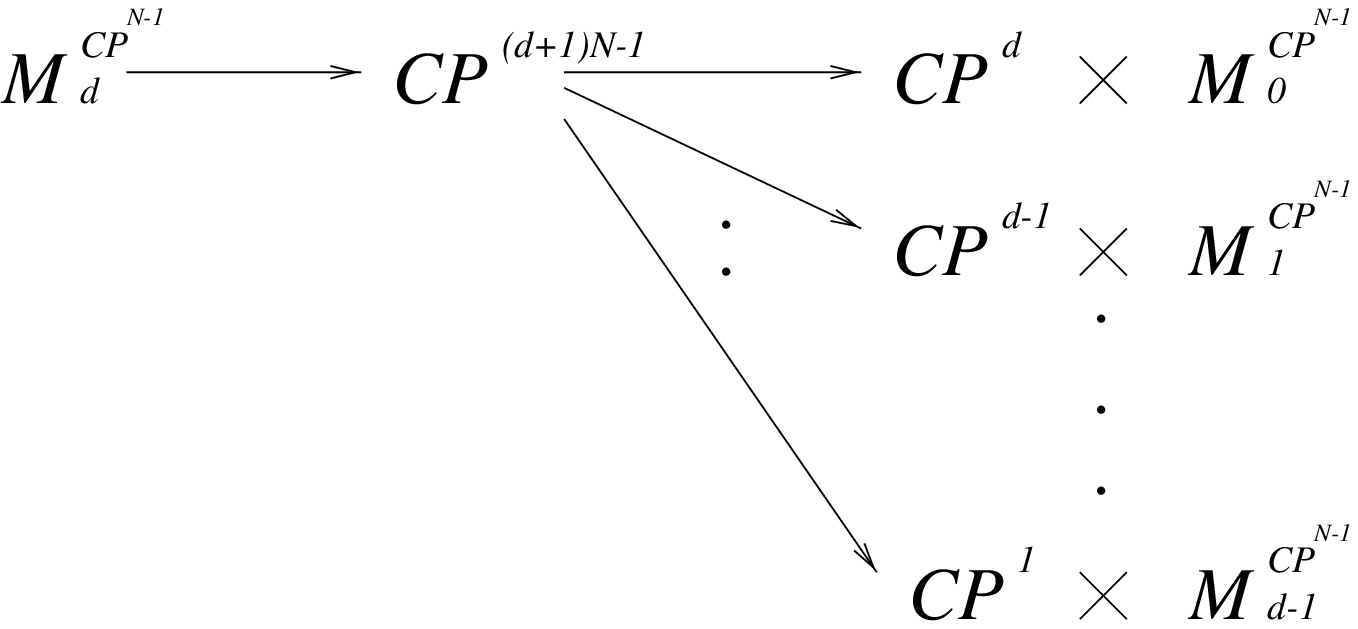}}
     \caption{\bf Set-Theoretic Short Exact Sequence}
\end{figure}
Since ${\cal M}_{0}^{CP^{N-1}}$ is isomorphic to $CP^{N-1}$, 
resolution of ${\cal M}_{1}^{CP^{N-1}}$ is given as: 
\begin{equation}
{\cal M}_{1}^{CP^{N-1}}\rightarrow CP^{2N-1} \rightarrow
CP^{1}\times CP^{N-1}.
\label{first}
\end{equation}
Now that we have the diagram of figure 3, what we have to do to
construct projective space resolution of ${\cal M}_{d}^{CP^{N-1}}$ is
to construct projective space resolution of $CP^{j}
\times{\cal M}_{d-j}^{CP^{N-1}},\;\;(j=1,2,\cdots,d)$. 
This can be done by applying $CP^{j}\times $ to each space in the  
resolution diagram of ${\cal M}_{d-j}^{CP^{N-1}}$. Then we can construct 
the resolution diagram of ${\cal M}_{d}^{CP^{N-1}}$ inductively.

From figure 1, we can see that each space $CP^{d_{1}}\times
CP^{d_{2}}\times \cdots \times CP^{d_{m}}\times CP^{N(d'+1)-1}$ induced from 
a sequence $(d_{1},d_{2},\cdots,d_{m};d')
\;\;\;(d_{j}\geq 1,\; d'\geq 0,\;\;\sum_{j=1}^{m}d_{j}+d'=d)$
 appear at least once and only once in the resolution diagram of 
${\cal M}_{d}^{CP^{N-1}}$.
Including the null sequence $(0;d)$ in the case of $CP^{(d+1)N-1}$,  
total number of such 
integer sequences is $1+\sum_{j=1}^{d}2^{j-1}=2^{d}$.

We also associate a pair 
$CP^{d_{1}}\times CP^{d_{2}}\times\cdots
\times CP^{d_{m}}\times CP^{(d'+1)N-1} 
\rightarrow CP^{d_{1}}\times CP^{d_{2}}\times\cdots
\times CP^{d_{m}}\times CP^{d'}\times CP^{N-1}$ with an 
integer sequence $(d_{1},d_{2},\cdots,d_{m}),\;\;(d_{j}\geq 1,\;\;
\sum_{j=1}^{m}d_{j}=k\;\;\; (k=1,\cdots,d-1))$. These pairs appear in the 
resolution diagram once and only once. If we formally associate 
the pair $CP^{(d+1)N-1}\rightarrow CP^{d}\times CP^{N-1}$ with $0$,
we can see that these pairs cover the diagram.  Neglecting 
the order of the integer sequence, we can take a partition 
$\sigma_{k}:k=d_{1}+d_{2}+\cdots+d_{m}\;\;(0\leq k\leq d-1)$ as the label 
of these pairs. Note that this label has multiplicity or, in other words, not 
one to one. We denote the multiplicity of $\sigma_{k}$ by $N(\sigma_{k})$ 
and  multiplicity of $j, \;\;(1\leq j\leq k)$, appearing in $\sigma_{k}$ by 
$mul(j,\sigma_{k})$.
Then we have 
\begin{equation}
N(\sigma_{k})=\frac{m!}{\prod_{j=1}^{k}mul(j,\sigma_{k})!}.
\label{parm}
\end{equation}
If we denote a set of partitions of integer $k$ by $P_{k}$, we see 
\begin{eqnarray}
&&\sum_{\sigma_{k}\in P_{k}}N(\sigma_{k})=2^{k-1}\;\;(k\geq 1),\no\\
&&\sum_{k=0}^{d-1}\sum_{\sigma_{k}\in P_{k}}N(\sigma_{k})=2^{d-1},
\label{exer}
\end{eqnarray}
which are easy exercises of elementary combinatorics.

As we mentioned in our previous paper \cite{jin}, these resolution structures 
do not occur in the quantum cohomology of degree $k$ hypersurfaces 
in $CP^{N-1}$ with $N-k\geq 2$ as we can see by dimensional counting,
but in the case of $N-k\leq 1$ and especially in the case of $N-k\leq 0$,
we will see the above structure appears. 
In these cases, we expect that a topological invariant obtained from 
integrating out the closed forms on ${\cal M}_{d}^{CP^{N-1}}$ can be
represented as an alternate sum of the contributions from each pair in the  
resolution diagram
(We denote length of partition $\sigma_{m}$ by $l(\sigma_{m})$):
\begin{equation}
(\mbox{top. inv. on ${\cal M}_{d}^{CP^{N-1}}$})=
\sum_{m=0}^{d-1}\sum_{\sigma_{m}\in P_{m}}(-1)^{l(\sigma_{m})}
N(\sigma_{m})(\mbox{contribution from one pair labeled by $\sigma_{m}$}).
\label{sisou}
\end{equation}
Here, we assumed that the contributions from each pair labeled by 
$\sigma_{m}$ are the same.  
In section 4, we will see the formula that can be regarded as an example of 
(\ref{sisou}). 
\section{Quantum K\"ahler Sub-Ring of Projective Hypersurfaces}
\subsection{Notation} 
In this section, we introduce the quantum K\"ahler sub-ring 
of the quantum cohomology ring of a degree $k$ hypersurface in
$CP^{N-1}$.
Let $M_{N}^{k}$ be a hypersurface of degree $k$ in $CP^{N-1}$.
 We denote by $QH^{*}_{e}(M_{N}^{k})$ the 
sub-ring of the quantum cohomology ring $QH^{*}(M_{N}^{k})$
generated by ${\cal O}_{e}$ induced from the K\"ahler form $e$ 
(or, equivalently the intersection $H\cap M_{N}^{k}$ between a hyperplane
class $H$ of $CP^{N-1}$ and $M_{N}^{k}$).
 The multiplication rule of $QH^{*}_{e}(M_{N}^{k})$ 
is determined by the Gromov-Witten invariant of genus $0$ 
$\langle {\cal O}_{e}{\cal O}_{{e}^{N-2-m}}
{\cal O}_{{e}^{m-1-(k-N)d}}\rangle_{d,M_{N}^{k}}$ and
it is given as follows:
\begin{eqnarray}
 L_{m}^{N,k,d} &:=&\frac{1}{k}\langle {\cal O}_{e}{\cal O}_{{e}^{N-2-m}}
{\cal O}_{{e}^{m-1-(k-N)d}}\rangle,\no\\
{\cal O}_{e}\cdot 1&=&{\cal O}_{e},\nonumber\\
{\cal O}_{e}\cdot{\cal O}_{{e}^{N-2-m}}&=&{\cal O}_{{e}^{N-1-m}}+
\sum_{d=1}^{\infty}L_{m}^{N,k,d}q^{d}{\cal O}_{{e}^{N-1-m+(k-N)d}},\no\\
q&:=&e^{t}.
\label{gm}
\end{eqnarray}
\begin{defi}
We call $L_{m}^{N,k,d}$ the structure constants of weighted degree $d$.
\end{defi}
Since $M_{N}^{k}$ is a complex $N-2$ dimensional manifold, we see that
the structure constants $L_{m}^{N,k,d}$
is non-zero only if the following condition is satisfied:
\begin{eqnarray}
L_{m}^{N,k,d}\neq 0&\Longrightarrow& 1\leq N-2-m\leq N-2, 
1\leq m-1+(N-k)d\leq N-2,\no\\
&\Longleftrightarrow &max.\{0,2-(N-k)d\}\leq m \leq min.\{N-3,N-1-(N-k)d\}.
\label{sel}
\end{eqnarray}
We rewrite (\ref{sel}) into 
\begin{eqnarray}
L_{m}^{N,k,d}\neq 0&\Longrightarrow& 0\leq m \leq (N-1)-(N-k)d\;\;(N-k\geq 2)
,\no\\
&\Longrightarrow& 1\leq m \leq N-3\;\;(N-k=1,d=1)
,\no\\
&\Longrightarrow& 0\leq m \leq N-1-(N-k)d\;\;(N-k=1,d\geq2)
,\no\\
&\Longrightarrow& 2+(k-N)d\leq m \leq N-3\;\;(N-k\leq 0).
\label{flasel}
\end{eqnarray}
From (\ref{flasel}), we easily see that the number of the non-zero
structure constants $L_{m}^{N,k,d}$ is finite except for the case of $N=k$.
Moreover, if $N\geq 2k$, the non-zero structure constants come only from
the $d=1$ part and the number of them is $k$, which is independent of $N$.
The $N\geq 2k$ region is studied by Beauville \cite{beauville}, 
and his result plays 
the role of an initial condition of our discussion later.
In the case of $N=k$, the multiplication rule of $QH^{*}_{e}(M_{k}^{k})$ is
given as follows:
\begin{eqnarray}
{\cal O}_{e}\cdot 1&=&{\cal O}_{e},\nonumber\\
{\cal O}_{e}\cdot{\cal O}_{{e}^{N-2-m}}&=&(1+\sum_{d=1}^{\infty}q^{d}L_{m}^{k,k,d}){\cal O}_{{e}^{N-1-m}}
\;\;(m=2,3,\cdots,N-3),\no\\
{\cal O}_{e}\cdot{\cal O}_{{e}^{N-3}}  &=&{\cal O}_{e^{N-2}}.
\label{calabi}
\end{eqnarray}
We introduce here the generating function of the structure constants 
of the Calabi-Yau hypersurface $M_{k}^{k}$:
\begin{equation}
L_{m}^{k,k}(e^{t}):=1+\sum_{d=1}^{\infty}L_{m}^{k,k,d}e^{dt}\;\;(m=2,\cdots 
k-3).
\end{equation}
  
 In this paper, we assume that $L_{m}^{N,k,d}$ is geometrically 
given by the following formula: 
\begin{eqnarray}
L_{m}^{N,k,d}&&=\frac{1}{k}\langle {\cal O}_{e}{\cal O}_{{e}^{N-2-m}}
{\cal O}_{{e}^{m-1-(k-N)d}}\rangle_{d,M_{N}^{k}}\no\\
&&=\frac{d}{k}\int_{\bar{\cal M}_{0,d,2}^{CP^{N-1}}}
c_{T}(\tilde{\pi}_{2}^{*}(R^{0}{\pi_{1}}_{*}(\phi_{1}^{*}(k\cdot H))))\wedge 
\phi_{1}^{*}(c_{1}^{N-2-n}(H))\wedge 
\phi_{2}^{*}(c_{1}^{n-1-(k-N)d}(H)).\no\\
\label{total}     
\end{eqnarray}
Here, $\bar{\cal M}_{0,d,n}^{CP^{N-1}}$ is the 
moduli space of the stable maps from 
$CP^{1}$ with $n$ points to $CP^{N-1}$ of  mapping degree $d$.
$\pi_{j}:\bar{\cal M}_{0,d,j}^{CP^{N-1}}\rightarrow 
\bar{\cal M}_{0,d,j-1}^{CP^{N-1}}$ is the forgetful map and
$\tilde{\pi}_{m}$ is given by
$\pi_{1}\circ\pi_{2}\circ\cdots\circ\pi_{m}:
\bar{\cal M}_{0,d,m}^{CP^{N-1}}\rightarrow\bar{\cal M}_{0,d,0}^{CP^{N-1}}$.
$\phi_{i}:\bar{\cal M}_{0,d,m}^{CP^{N-1}}\rightarrow 
CP^{N-1}\;\;(j=1,2,\cdots,m)$ is the evaluation map at the $j$-th puncture.
The direct image sheaf $R^{0}{\pi_{1}}_{*}(\phi_{1}^{*}(k\cdot H))$ is 
considered in the setting of the fibration $CP^{1}\rightarrow 
\bar{\cal M}_{0,d,1}^{CP^{N-1}}\rightarrow\bar{\cal
M}_{0,d,0}^{CP^{N-1}}$.
The r.h.s. of (\ref{total}) can be evaluated by the toroidal calculation 
method of Kontsevich \cite{tor}, \cite{j} and we will use this 
data to check numerically our prediction formula 
in section 5.
\subsection{Review of Results for Fano and Calabi-Yau Hypersurfaces
and Virtual Structure Constants}
Let us summarize the results of \cite{cj}. In \cite{cj}, 
we showed that the structure constants $L_{m}^{N,k,d}$ 
of $QH_{e}^{*}(M_{N}^{k}),\:\:(N-k\geq 2)$, can be obtained by 
applying the recursive formulas in Appendix B, that represents
$L_{m}^{N,k,d}$ in terms of $L_{m'}^{N+1,k,d'},\;\;(d'\leq d)$,
 with the initial 
condition $L_{m}^{N,k,1}$ (they are non-zero only for
$m=0,1,2,\cdots,k-1$ and do not depend on $N$) 
and $L_{m}^{N,k,d}=0,\;\;(d\geq 2)$, in the $N\geq 2k$ region.
These recursive formulas naturally lead us to the relation:
\begin{equation}
({\cal O}_{e})^{N-1}-k^{k}({\cal O}_{e})^{k-1}q=0
\label{jincolli}
\end{equation}
of $QH_{e}^{*}(M_{N}^{k})\:\:(N-k\geq 2)$            
by a descending induction from the result of Beauville.
 This formula shows that the resolution structure in section 2 
do not occur in the case of $N-k\geq 2$. 

In the $N-k=1$ case, the recursive formulas receive modification 
only in the $d=1$ part:
\begin{equation}
L_{m}^{k+1,k,1}=L_{m}^{k+2,k,1}-L_{0}^{k+2,k,1}=L_{m}^{k+2,k,1}-k!.
\label{givgiv}
\end{equation}
This leads us to the following relation of $QH_{e}^{*}(M_{k+1}^{k})$:
\begin{equation}
({\cal O}_{e}+k!q)^{N-1}-k^{k}({\cal O}_{e}+k!q)^{k-1}q=0.
\label{ch1}
\end{equation} 
 
The structure constant $L_{m}^{k,k,d}$ for a Calabi-Yau hypersurface 
do not obey the recursive formulas in appendix B.    
We introduce here virtual structure constants $\tilde{L}_{m}^{N,k,d}$
as follows.
\begin{defi}
Let $\tilde{L}_{m}^{N,k,d}$ be the rational number obtained by applying
the recursion relations of Fano hypersurfaces in Appendix B
 for arbitrary $N$ and $k$ with the initial condition
 $L_{n}^{N,k,1},\;\;(N\geq 2k)$, and 
$L_{n}^{N,k,d}=0,\;\;(d\geq 2,\;\;N\geq 2k)$. 
\end{defi}
{\bf Remark 1}
In the $N-k\geq2$ region, $\tilde{L}_{m}^{N,k,d}=L_{m}^{N,k,d}$.

We define the generating function of the virtual structure constants of 
the Calabi-Yau hypersurface $M_{k}^{k}$
as follows:
\begin{eqnarray}
\tilde{L}^{k,k}_{n}(e^{x})&:=&1+
\sum_{d=1}^{\infty}\tilde{L}^{k,k,d}_{n}e^{dx},\nonumber\\
&&(n=0,1,\cdots,k-1).
\label{str2}
\end{eqnarray}
Then we conjectured that $\tilde{L}^{k,k}_{n}(e^{x})$ gives us the 
information of the B-model of the mirror manifold of $M_{k}^{k}$. 
More explicitly, we conjectured 
\begin{equation}
\tilde{L}_{0}^{k,k}(e^{x})=
\sum_{d=0}^{\infty}\frac{(kd)!}{(d!)^{k}}e^{dx},
\label{po}
\end{equation}
where the r.h.s. of (\ref{po}) is the power series solution of 
the ODE for the period integral of the mirror manifold of $M_{k}^{k}$,
\begin{equation}
((\frac{d}{dx})^{k-1}-k^{k}e^{x}(\frac{d}{dx}+\frac{1}{k}) 
(\frac{d}{dx}+\frac{2}{k})\cdots(\frac{d}{dx}+\frac{k-1}{k}))w(x)=0.
\label{ode1}
\end{equation}
Moreover we can obtain the mirror map $t=t(x)$ without 
using the mirror conjecture:
\begin{equation}
t(x)=x+\int_{-\infty}^{x}dx'({\tilde{L}^{k,k}_{1}(e^{x'})}-1)
=x+\sum_{d=1}^{\infty}\frac{\tilde{L}^{k,k,d}_{1}}{d}e^{dx}.
\end{equation}
With these conjectures, we can construct 
the mirror transformation that transforms 
the virtual structure constants of the  Calabi-Yau hypersurface into 
the real ones as follows:
\begin{equation}
L^{k,k}_{n}(e^{t})=\frac{\tilde{L}^{k,k}_{n}(e^{x(t)})}
{\tilde{L}^{k,k}_{1}(e^{x(t)})}.
\label{mt}
\end{equation}

Here we will discuss whether we can obtain useful informations 
from the virtual structure constants $\tilde{L}_{m}^{N,k,d}$ in the 
$N<k$ region. 
Note that $\tilde{L}_{m}^{N,k,d}$ is non-zero only if $0\leq m\leq N-1+(k-N)d$.
So a hypersurface with non-positive first Chern class has infinite number
of $\tilde{L}_{m}^{N,k,d}$, though only a finite number of
$L_{m}^{N,k,d}$ appear in $QH^{*}_{e}(M_{N}^{k}),\;(N<k)$. 
In this case, the recursive formulas in Appendix B restricted to 
$\tilde{L}_{0}^{N,k,d}$ yield 
the following recursive relations:
\begin{equation}
\tilde{L}_{0}^{N,k,d}=\frac{d!}{d^{d}}\tilde{L}_{0}^{N+1,k,d}.
\end{equation}
Then we obtain from (\ref{po})  
\begin{equation}
1+\sum_{d=1}^{\infty}\tilde{L}_{0}^{N,k,d}e^{dx}
=\sum_{d=0}^{\infty}\frac{(kd)!}{(d!)^{N}\cdot (d^{d})^{k-N}}e^{dx}.
\label{neg}
\end{equation}
The r.h.s. of (\ref{neg}) resembles the power series solution
of the ODE 
\begin{equation}
((\frac{d}{dx})^{N-1}-k^{k}e^{x}(\frac{d}{dx}+\frac{1}{k}) 
(\frac{d}{dx}+\frac{2}{k})\cdots(\frac{d}{dx}+\frac{k-1}{k}))w(x)=0,
\label{ode2}
\end{equation}
given by  
\begin{equation}
w(x)=\sum_{d=0}^{\infty}\frac{(kd)!}{(d!)^{N}}e^{dx}.
\label{giv}
\end{equation}
These formulas suggest the existence of a 
mirror manifold of $M_{N}^{k},\;(N<k)$. 
So we proceed further expecting that there exists formulas that 
transforms the virtual structure constants into the real structure
constants like the case of $M_{k}^{k}$.
However, these information 
are less important in the relation with the structure constants of
the real quantum cohomology ring $L_{m}^{N,k,d}$ if we reconsider
analogy with the case of Calabi-Yau hypersurfaces. In the case of 
Calabi-Yau hypersurfaces, the information of 
$\tilde{L}_{m}^{k,k,d},\;(m=1,2,\cdots,k-2)$, is used to construct 
$L_{m}^{k,k,d},\;(m=2,\cdots,k-3)$, and  
$\tilde{L}_{0}^{k,k,d}=\tilde{L}_{k-1}^{k,k,d}$ are thrown away. 
That corresponds to 
the ``trivialization of line bundle factor''. One of the reasons why
we need $\tilde{L}_{1}^{k,k,d}=\tilde{L}_{k-2}^{k,k,d}$ comes 
from the realization of the ``flat metric condition'', or more formally, 
we need cancellation of the formulas that express $L_{m}^{k,k,d}$ 
in terms of $\tilde{L}_{n}^{k,k,d'},\;(d'\leq d)$, 
if we set $m=1$ or $m=k-2$. This condition assures us that the identity 
operator in $QH_{e}^{*}(M_{k}^{k})$ doesn't receive quantum correction.
Similarly, we require the formulas 
that express $L_{m}^{N,k,d},\; (N<k)$, in terms of 
$\tilde{L}_{n}^{N,k,d'},\;(d'\leq d)$, to cancel if we formally 
set $m=1+(k-N)d$ or $m=N-2$. Moreover, we assume that 
what we need in constructing $L_{m}^{N,k,d}$, which is non-zero only if 
$2+(k-N)d\leq m \leq N-3$, is $\tilde{L}_{n}^{N,k,d'},\;(d'\leq
d,\;1+(k-N)d' \leq n \leq N-2)$. In other words, we threw away infinite 
number of $\tilde{L}_{n}^{N,k,d}$ in the $N<k$ case except for 
$\tilde{L}_{n}^{N,k,d},\;(1+(k-N)d \leq n \leq N-2)$. 
Later on, we call the formulas that express $L_{m}^{N,k,d}$ in terms 
of $\tilde{L}_{n}^{N,k,d'},\;(d'\leq d)$, the ``generalized mirror
transformation''.

 We introduce the following notation.
\begin{defi}
Let $V^{N,k,d}$ (resp. $\tilde{V}^{N,k,d}$) be the vector space of weighted
homogeneous polynomial of $L_{n}^{N,k,d'}\;\;(d'\leq d)$ 
(resp. $\tilde{L}_{n}^{N,k,d'}\;\;(d'\leq d)$) of degree $d$.
\end{defi}
Then consider the commutative diagram
\[\begin{array}{ccccccccc}
\cdots&\mapleft{\tilde{\phi}_{N+2}}&\tilde{V}^{N+2,k,d}&
\mapleft{\tilde{\phi}_{N+1}} 
&\tilde{V}^{N+1,k,d}&\mapleft{\tilde{\phi}_{N}} 
&\tilde{V}^{N,k,d}&\mapleft{\tilde{\phi}_{N-1}}&\cdots\\
&&\mapup{m_{N+2}^{k}}
&&\mapup{m_{N+1}^{k}}&&\mapup{m_{N}^{k}}&&\\
\cdots&\mapleft{\phi_{N+2}}&V^{N+2,k,d}&
\mapleft{\phi_{N+1}}&V^{N+1,k,d}&\mapleft{\phi_{N}} 
&V^{N,k,d}&\mapleft{\phi_{N-1}}&\cdots .\\
\end{array}\]
The map $\phi_{N}$ is the recursive formula between $L_{n}^{N+1,k,d}$
and $L_{n}^{N,k,d}$
, which is difficult to calculate for the $d\geq 3$ cases  
 (we will construct them for the $d=1,2$ cases in the next section) 
, and $\tilde{\phi}_{N}$, which is 
the recursive formula between $\tilde{L}_{n}^{N+1,k,d}$
and $\tilde{L}_{n}^{N,k,d}$, is obvious by definition.
We denote by $m_{N}^{k}$  the generalized mirror transformation (of
course, $m_{N}^{k}=id$ if $N-k\geq 2$).
Then if we can determine $m_{N}^{k}$, we can construct 
 $\phi_{N}$ from equation $\phi_{N}=(m_{N+1}^{k})^{-1}
\circ\tilde{\phi}_{N}\circ(m_{N}^{k})$. Conversely, 
we can construct $m_{N}^{k}$ if we can construct $\phi_{N}$.
In the next section, we will give some results as some clues for 
constructing the generalized mirror transformation.

\section{Some Preparatory Results}
\subsection{Recursive Formula for the $d=1,2$ cases}
In this subsection, we prove a theorem on the structure constants of 
$QH^{*}_{e}(M_{N}^{k}),\;\;(k>N)$, coming from the quantum correction
of the $d=1,2$ part. Unfortunately, our method used in the proof of the
theorem is effective only in these cases for the   
general type hypersurfaces. However, the result gives us a hint for the 
construction of the generalized mirror transformation proposed in section 5.
 \begin{theorem}
In the case of $k-N\geq0$, the following formulas  
expressing $L_{n}^{N,k,d}$ in terms of $L_{n}^{N+1,k,d'}\;\;(d'\leq d)$ 
for the $d=1,2$ cases hold: 
\begin{eqnarray}
L_{n}^{N,k,1}&=&L_{n}^{N+1,k,1}-L_{1+(k-N)}^{N+1,k,1},\nonumber\\
L_{n}^{N,k,2}&=&
\frac{1}{2}(L_{n}^{N+1,k,2}+L_{n-1}^{N+1,k,2}+
2L_{n}^{N+1,k,1}L_{n-(k-N)}^{N+1,k,1})\no\\
&&-\frac{1}{2}(L_{1+2(k-N)}^{N+1,k,2}+L_{2(k-N)}^{N+1,k,2}+
2L_{1+2(k-N)}^{N+1,k,1}L_{1+(k-N)}^{N+1,k,1})\nonumber\\
&&-2L_{1+(k-N)}^{N+1,k,1}(\sum_{j=0}^{k-N}(L_{n-j}^{N+1,k,1}
-L_{1+2(k-N)-j}^{N+1,k,1})).
\label{conic}
\end{eqnarray}
\end{theorem}
{\it proof)}
We roughly show a sketch of the proof along the line of \cite{cj}.
Set
\begin{equation}
[A_{a_{1}},A_{a_{2}},\cdots,A_{a_{n}};N,k,d]:=\langle{\cal O}_{e^{a_{1}}}{\cal O}_{e^{a_{2}}}\cdots {\cal
O}_{e^{a_{n}}}\rangle_{d,M_{N}^{k}}
\label{linear}
\end{equation}
where $A_{a_{j}}$ is a linear subspace of codimension $a_{j}$ in
$CP^{N-1}$, and we assume $A_{a_{j}}$'s are in general position.  
Let 
\begin{equation}
G[A_{a_{1}}\cap H ,A_{a_{2}}\cap H,\cdots,A_{a_{k}}\cap H,\cdots 
A_{a_{k+1}},\cdots,A_{a_{k+m}};N,k,d] 
\label{special}
\end{equation}
be the correlation function in special position,
where $H$ is fixed hyperplane in $CP^{N-1}$, so the linear subspaces 
$A_{a_{1}}\cap H ,\cdots, A_{a_{k}}\cap H$ are not in general position 
and $\cap_{j=1}^{k}(A_{a_{j}}\cap H)=(\cap_{j=1}^{k}A_{a_{j}})\cap H$.
Then $M_{N}^{k}$ is embedded in $M_{N+1}^{k}$ as
$M_{N+1}^{k}\cap H$. Note that an irreducible rational curve of degree $d$
intersecting fixed hyperplane $H$ with $d+1$ points must lie on $H$.
Then, we have the equation 
\begin{equation}
G[A_{a_{1}}\cap H ,A_{a_{2}}\cap H,\cdots,A_{a_{d+1}}\cap H ;N+1,k,d]
=[A_{a_{1}},A_{a_{2}},\cdots,A_{a_{d+1}};N,k,d]+R
\label{d1}
\end{equation}
where $R$ consists of the contribution from the reduced curves with one 
component
lying on $M_{N}^{k}$ and the other components lying on $M_{N+1}^{k}$, and of  
the contribution from some excess intersection that appear 
in the $d\geq 3$ case that is hard to 
determine geometrically. Obviously, $R$ is $0$ if $d$ equals $1$, and in 
the case of $d=2$, 
the contribution from the reduced curves are given by the formula 
\begin{equation}
R=\frac{1}{k}[A_{N-2-(k-N)};N+1,k,1]\cdot
[A_{1+(k-N)},A_{N-2-n},A_{n-1-2(k-N)},A_{1};N,k,1].
\label{select}
\end{equation}
The correlation function
$G[A_{a_{1}}\cap H ,A_{a_{2}}\cap H,\cdots,A_{a_{d+1}}\cap H ;N+1,k,d]$ 
can be evaluated by the specialization formula 
\begin{eqnarray}
&& G[A_{a_{1}}\cap H , \cdots,
A_{a_{s}}\cap H,
A_{a_{s+1}+1},A_{a_{s+2}+1},
\cdots, A_{a_{s+t}+1};d,N+1,k]\nonumber\\
&=& G[A_{a_{1}}\cap H , \cdots,
A_{a_{s}}\cap H,
A_{a_{s+1}}\cap H ,A_{a_{s+2}+1},\cdots,
A_{a_{s+t}+1};d,N+1,k]+\nonumber\\
&&\sum_{j=1}^{s} G[A_{a_{1}}\cap H ,
\cdots,A_{a_{j-1}}\cap H,A_{a_{j}+a_{s+1}}\cap H ,
A_{a_{j+1}}\cap H ,
\cdots,A_{a_{s}}\cap H,\nonumber\\
&& A_{a_{s+2}+1},\cdots,
A_{a_{s+t}+1};d,N+1,k].
\label{spec}
\end{eqnarray}
We apply the specialization formula to 
$G[A_{N-2-n}\cap H ,A_{n-1-(k-N)}\cap H;1,N+1,k]$ and to 
$G[A_{N-2-n}\cap H ,A_{n-1-2(k-N)}\cap H,A_{1}\cap H;2,N+1,k]$,
and obtain 
\begin{eqnarray}
&&[A_{N-2-n},A_{n-1-(k-N)};1,N+1,k]\no\\
&&=G[A_{N-2-n}\cap H ,A_{n-1-(k-N)}\cap H;1,N+1,k]\no\\
&&=[A_{N-1-n},A_{n-(k-N)};1,N+1,k]-[A_{N-2-(k-N)};1,N+1,k]\no\\
&&[A_{N-2-n},A_{n-1-2(k-N)},A_{1};2,N+1,k]+R\no\\
&&=G[A_{N-2-n}\cap H ,A_{n-1-2(k-N)}\cap H,A_{1}\cap H;2,N+1,k]\no\\
&&=[A_{N-1-n},A_{n-2(k-N)},A_{2};2,N+1,k]\no\\
&&-[A_{N-n},A_{n-2(k-N)};2,N+1,k]-[A_{N-1-n},A_{n+1-2(k-N)};2,N+1,k]\no\\
&&-[A_{N-2-2(k-N)},A_{2};2,N+1,k]+2[A_{N-1-2(k-N)};2,N+1,k].
\label{apply}
\end{eqnarray}
Then using the microscopic version of the associativity equation \cite{dub}, 
\cite{km}, 
we rewrite the Gromov-Witten invariant of $M_{N+1}^{k}$ in 
terms of $L_{n}^{N+1,k,1}$ and $L_{n}^{N+1,k,2}$
\begin{eqnarray}
L_{n}^{N,k,1}&=&L_{n}^{N+1,k,1}-L_{1+(k-N)}^{N+1,k,1},\nonumber\\
L_{n}^{N,k,2}+\frac{1}{k}R&=&
\frac{1}{2}(L_{n}^{N+1,k,2}+L_{n-1}^{N+1,k,2}-L_{1+2(k-N)}^{N+1,k,2}
-L_{2(k-N)}^{N+1,k,2})\nonumber\\
&&+L_{n}^{N+1,k,1}(L_{n-(k-N)}^{N+1,k,1}-L_{1+(k-N)}^{N+1,k,1})\nonumber\\
&&-L_{1+(k-N)}^{N+1,k,1}(\sum_{j=1}^{k-N}(L_{n+j-1-(k-N)}^{N+1,k,1}
-L_{j+(k-N)}^{N+1,k,1})).
\label{ct}
\end{eqnarray}
The first line of (\ref{ct}) directly leads to the theorem 
for the $d=1$ case.
By the associativity formula and the first line of 
(\ref{ct}), the second line of (\ref{ct})
turns out to be the desired formula for the $d=2$ case. Q.E.D.

\begin{cor}
The generalized mirror transformation for  $d=1,2$ is given by:
\begin{eqnarray}
L_{n}^{N,k,1}&=&\tilde{L}_{n}^{N,k,1}-\tilde{L}_{1+(k-N)}^{N,k,1},\nonumber\\
L_{n}^{N,k,2}&=&\tilde{L}_{n}^{N,k,2}-\tilde{L}_{1+2(k-N)}^{N,k,2}
-2\tilde{L}_{1+(k-N)}^{N,k,1}(\sum_{j=0}^{k-N}(\tilde{L}_{n-j}^{N,k,1} 
- \tilde{L}_{1+2(k-N)-j}^{N,k,1})).
\label{mirr}
\end{eqnarray}
\end{cor}
\subsection{Mirror Transformation of Calabi-Yau Hypersurface 
and Schur Polynomials}
Here, we show some explicit results on the mirror transformation on the 
Calabi-Yau hypersurfaces as another clue for constructing 
the generalized mirror transformation. 
In the next section, we propose some conjectures on the structure
of the generalized mirror transformation on the quantum cohomology 
of the general type ($N-k < 0$) hypersurfaces.
First, we expand (\ref{mt}) into a power series in $e^{t}$ (\ref{mt}) and 
obtain the following
formulas up to degree $4$.
\begin{eqnarray}
L^{k,k,1}_{n}&=&\tilde{L}^{k,k,1}_{n}-\tilde{L}^{k,k,1}_{1},\nonumber\\
L^{k,k,2}_{n}&=&\tilde{L}^{k,k,2}_{n}-\tilde{L}^{k,k,2}_{1}
                -2\tilde{L}^{k,k,1}_{1}
                  (\tilde{L}^{k,k,1}_{n}-\tilde{L}^{k,k,1}_{1}),\nonumber\\
L^{k,k,3}_{n}&=&\tilde{L}^{k,k,3}_{n}-\tilde{L}^{k,k,3}_{1}
                -3\tilde{L}^{k,k,1}_{1}
                  (\tilde{L}^{k,k,2}_{n}-\tilde{L}^{k,k,2}_{1})+
                  (\frac{9}{2}(\tilde{L}^{k,k,1}_{1})^{2}-
                   \frac{3}{2}\tilde{L}^{k,k,2}_{1})
                  (\tilde{L}^{k,k,1}_{n}-\tilde{L}^{k,k,1}_{1}),\nonumber\\
L^{k,k,4}_{n}&=&\tilde{L}^{k,k,4}_{n}-\tilde{L}^{k,k,4}_{1}
                -4\tilde{L}^{k,k,1}_{1}
                  (\tilde{L}^{k,k,3}_{n}-\tilde{L}^{k,k,3}_{1})+
                  (8(\tilde{L}^{k,k,1}_{1})^{2}-
                   2\tilde{L}^{k,k,2}_{1})
                  (\tilde{L}^{k,k,2}_{n}-\tilde{L}^{k,k,2}_{1})\nonumber\\
                  && +(-\frac{32}{3}(\tilde{L}^{k,k,1}_{1})^{3}
                   +8\tilde{L}^{k,k,2}_{1}\tilde{L}^{k,k,1}_{1}-        
                    \frac{4}{3}\tilde{L}^{k,k,3}_{1})
                  (\tilde{L}^{k,k,1}_{n}-\tilde{L}^{k,k,1}_{1}).
\label{mt2}
\end{eqnarray}
Looking at (\ref{mirr}) and (\ref{mt2}), we speculate a close 
resemblance between these two formulas. We can imagine that 
(\ref{mirr}) can be obtained from (\ref{mt2}) by substituting 
$\tilde{L}^{N,k,d}_{1+(k-N)d}$ into 
$\tilde{L}^{k,k,d}_{1}$ and 
$\sum_{j=0}^{(d-d')(k-N)}(\tilde{L}_{n-j}^{N,k,d'} 
- \tilde{L}_{1+(k-N)d-j}^{N,k,d'})$ into
$\tilde{L}^{k,k,d'}_{n}-\tilde{L}^{k,k,d'}_{1}$.
This speculation is not exactly correct, but not far from the truth.
Before going ahead, we introduce a more general 
result on (\ref{mt}).   
\begin{prop}
Let $P_{m}$ be the set of partitions of $m$ into positive 
integers and $\sigma_{m}$ be an element of $P_{m}$. We also denote 
the length of a partition $\sigma_{m}$ by $l(\sigma_{m})$ (i.e., 
$\sigma_{m}:m=d_{1}+d_{2}+
\cdots+d_{l(\sigma_{m})},\;\;d_{1}\geq d_{2}\geq\cdots
\geq d_{l(\sigma_{m})}\geq1$).   
Then the above mirror transformation takes the form
\begin{equation}
L^{k,k,d}_{n}=\sum_{m=0}^{d-1}\sum_{\sigma_{m}\in P_{m}}
(C_{m}^{d}(\sigma_{m})\cdot\prod_{i=1}^{l(\sigma_{m})}
(\tilde{L}_{1}^{k,k,d_{i}}))\cdot
(\tilde{L}^{k,k,d-m}_{n}-\tilde{L}^{k,k,d-m}_{1}),
\label{gene}
\end{equation}
where $C_{m}^{d}(\sigma_{m})$ is a constant coefficient.
\end{prop}
{\it proof )}
For brevity, we introduce the notation 
$\tilde{L}_{n}^{k.k}(z)=\sum_{d=0}^{\infty}a_{d}z^{d}$, 
$\tilde{L}_{1}^{k.k}(z)=\sum_{d=0}^{\infty}b_{d}z^{d}$ and 
$1/(\tilde{L}_{1}^{k.k}(z))=\sum_{d=0}^{\infty}c_{d}z^{d}$.
Then we can easily see the following identity,
\begin{equation}
\frac{\tilde{L}_{n}^{k.k}(z)}{\tilde{L}_{1}^{k.k}(z)}
=1+\sum_{d=1}^{\infty}(\sum_{j=1}^{d}c_{d-j}(a_{j}-b_{j}))z^{d}.
\label{frac}
\end{equation} 
 Proposition immediately follows from this identity. Q.E.D.

We propose here the following theorem, which suggests a close relation 
between the mirror transformation for Calabi-Yau hypersurfaces (\ref{mt})
and the elementary Schur polynomials associated with 
the partitions of integers, or Young diagrams.
\begin{theorem}
The mirror transformation is explicitly written as 
\begin{equation}
L^{k,k,d}_{n}=\sum_{m=0}^{d-1}Res_{z=0}(z^{-m-1}
\exp(-d\cdot(\sum_{j=1}^{\infty}\frac{\tilde{L}_{1}^{k,k,j}}{j}z^{j})))
\cdot(\tilde{L}^{k,k,d-m}_{n}-\tilde{L}^{k,k,d-m}_{1}).
\label{schur}
\end{equation}
\end{theorem}
The proof will be given in Appendix A.
\begin{cor}
\begin{equation}
C_{m}^{d}(\sigma_{m})=
(-1)^{l(\sigma_{m})}\frac{d^{l(\sigma_{m})}}
{\prod_{j=1}^{l(\sigma_{m})}d_{j} \prod_{i=1}^{m}mul(i,\sigma_{m})!}
=(-1)^{l(\sigma_{m})}\frac{d^{l(\sigma_{m})}\cdot N(\sigma_{m})}
{\prod_{j=1}^{l(\sigma_{m})}d_{j}\cdot l(\sigma_{m})!},
\label{corori}
\end{equation}
where $mul(i,\sigma_{m})$ is the multiplicity of $i$ in the partition 
$\sigma_{m}\in P_{m}$ and $N(\sigma_{m})$ is the number of the 
distinct elements 
in  the orbit of the natural action of symmetric group $S_{l(\sigma_{m})}$
on the ordered sequence $(d_{1},d_{2},\cdots,d_{l(\sigma_{m})})$. 
\end{cor}
Note that the multiplicity $N(\sigma_{m})$ 
of the partition $\sigma_{m}$ in section 2 
appears in $C_{m}^{d}(\sigma_{m})$.
(\ref{gene}) and (\ref{corori}) can be regarded as an example of 
(\ref{sisou}). More explicitly, we expect the following 
correspondence: 
\begin{eqnarray}
&&(\mbox{the contribution from one pair labeled by $\sigma_{m}$ in the 
resolution diagram of ${\cal M}_{d}^{CP^{N-1}}$})\no\\
&&=\frac{d^{l(\sigma_{m})}}{(\prod_{j=1}^{l(\sigma_{m})}d_{j})l(\sigma_{m})!}
(\prod_{j=1}^{l(\sigma_{m})}\tilde{L}_{1}^{N,k,d_{j}})
(\tilde{L}_{n}^{N,k,d-m}-\tilde{L}_{1}^{N,k,d-m}).
\label{spospo}
\end{eqnarray}

We also introduce here a notation on partitions of integers,
which is to be used in the next section and in the proof of Theorem 2. 
\begin{defi}
For two partitions $\sigma_{m}\in P_{m}$ and $\sigma_{n}\in P_{n}$,
we define $\sigma_{m}\cup \sigma_{n}\in P_{m+n}$ to be 
the partition whose parts are 
those of $\sigma_{m}$ and $\sigma_{n}$, arranged in descending order.
\end{defi}

\section{Search for Generalized Mirror Transformation}
\subsection{General Conjecture}
Now, we propose a conjecture on the generalized mirror
transformation, bearing in mind the speculation in the previous
subsection. 
\begin{conj}
The map $m_{N}^{k}$ is given as 
\begin{equation}
L^{N,k,d}_{n}=\sum_{m=0}^{d-1}\sum_{\sigma_{m}\in P_{m}}
(C_{m}^{d}(\sigma_{m})\cdot\prod_{i=1}^{l(\sigma_{m})}
(\tilde{L}_{1+(k-N)d_{i}}^{N,k,d_{i}}))\cdot
G_{d-m}^{N,k,d}(n;\sigma_{m}),
\label{gene2}
\end{equation}
where $G_{d-m}^{N,k,d}(n;\sigma_{m})$ is
a degree $d-m$ weighted 
homogeneous polynomial of $\tilde{L}^{N,k,d}_{n}$.
\end{conj}
Here $G_{d-m}^{N,k,d}(n;\sigma_{m})$ is yet to be determined.
$G_{d-m}^{N,k,d}(n;\sigma_{m})$ must satisfy the following conditions.\\
(i) flat metric condition
\begin{equation}
G_{d-m}^{N,k,d}(1+(k-N)d;\sigma_{m})=G_{d-m}^{N,k,d}(N-2;\sigma_{m})=0.
\end{equation}
(ii) symmetry
\begin{equation}
G_{d-m}^{N,k,d}(n;\sigma_{m})=G_{d-m}^{N,k,d}(N-1+(k-N)d-n;\sigma_{m}).
\end{equation}
Moreover, we assume, \\
(iii) $G_{d-m}^{N,k,d}(n;\sigma_{m})$ consists of
$\tilde{L}_{l}^{N,k,d'}$'s
that satisfy the condition,
\begin{eqnarray} 
&&\tilde{L}_{n-j}^{N,k,d'},\;(1\leq d'\leq d),\;(0\leq j \leq
(k-N)(d-d')), \no\\
&&\tilde{L}_{1+(k-N)d-j}^{N,k,d'},
\;(1\leq d'\leq d),\;(0\leq j \leq (k-N)(d-d')). 
\end{eqnarray}
Under these assumptions, we found the following numerical 
results for some lower 
$k$'s, using (\ref{total}).
\begin{eqnarray}
L_{5}^{k-1,k,3}&=&\tilde{L}_{5}^{k-1,k,3}-\tilde{L}_{4}^{k-1,k,3}-
3\tilde{L}_{2}^{k-1,k,1}(\tilde{L}_{5}^{k-1,k,2}-\tilde{L}_{3}^{k-1,k,2})
\nonumber\\
&&-\frac{3}{2}\tilde{L}_{3}^{k-1,k,2}(\tilde{L}_{5}^{k-1,k,1}-
\tilde{L}_{2}^{k-1,k,1})\nonumber\\
&&+\frac{9}{2}(\tilde{L}_{2}^{k-1,k,1})^{2}(\tilde{L}_{5}^{k-1,k,1}+
\tilde{L}_{4}^{k-1,k,1}-\tilde{L}_{3}^{k-1,k,1}-\tilde{L}_{2}^{k-1,k,1}),
\label{ex1}
\end{eqnarray}
\begin{eqnarray}
L_{6}^{k-1,k,3}&=&\tilde{L}_{6}^{k-1,k,3}-\tilde{L}_{4}^{k-1,k,3}-
3\tilde{L}_{2}^{k-1,k,1}(\tilde{L}_{6}^{k-1,k,2}+\tilde{L}_{5}^{k-1,k,2}
-\tilde{L}_{4}^{k-1,k,2}-\tilde{L}_{3}^{k-1,k,2}\nonumber\\
&&+\tilde{L}_{6}^{k-1,k,1}\tilde{L}_{4}^{k-1,k,1}-
\tilde{L}_{6}^{k-1,k,1}\tilde{L}_{3}^{k-1,k,1}-
\tilde{L}_{5}^{k-1,k,1}\tilde{L}_{3}^{k-1,k,1}\nonumber\\
&&+\tilde{L}_{5}^{k-1,k,1}\tilde{L}_{2}^{k-1,k,1}-
\tilde{L}_{4}^{k-1,k,1}\tilde{L}_{2}^{k-1,k,1}+
\tilde{L}_{3}^{k-1,k,1}\tilde{L}_{3}^{k-1,k,1})
\nonumber\\
&&-\frac{3}{2}\tilde{L}_{3}^{k-1,k,2}(\tilde{L}_{6}^{k-1,k,1}+
\tilde{L}_{5}^{k-1,k,1}-\tilde{L}_{3}^{k-1,k,1}-
\tilde{L}_{2}^{k-1,k,1})\no\\
&&+\frac{9}{2}(\tilde{L}_{2}^{k-1,k,1})^{2}(\tilde{L}_{6}^{k-1,k,1}+2
\tilde{L}_{5}^{k-1,k,1}-2\tilde{L}_{3}^{k-1,k,1}-\tilde{L}_{2}^{k-1,k,1}),
\label{ex2}
\end{eqnarray}
\begin{eqnarray}
L_{6}^{k-1,k,4}&=&\tilde{L}_{6}^{k-1,k,4}-\tilde{L}_{5}^{k-1,k,4}
-4\tilde{L}_{2}^{k-1,k,1}(\tilde{L}_{6}^{k-1,k,3}-\tilde{L}_{4}^{k-1,k,3})
\nonumber\\
&&-2\tilde{L}_{3}^{k-1,k,2}(\tilde{L}_{6}^{k-1,k,2}-\tilde{L}_{3}^{k-1,k,2})
-\frac{4}{3}\tilde{L}_{4}^{k-1,k,3}(\tilde{L}_{6}^{k-1,k,1}-
\tilde{L}_{2}^{k-1,k,1})\nonumber\\
&&+8(\tilde{L}_{2}^{k-1,k,1})^2(\tilde{L}_{6}^{k-1,k,2}+\tilde{L}_{5}^{k-1,k,2}
-\tilde{L}_{4}^{k-1,k,2}-\tilde{L}_{3}^{k-1,k,2}\nonumber\\
&&+\tilde{L}_{6}^{k-1,k,1}\tilde{L}_{4}^{k-1,k,1}-
\tilde{L}_{6}^{k-1,k,1}\tilde{L}_{3}^{k-1,k,1}-
\tilde{L}_{5}^{k-1,k,1}\tilde{L}_{3}^{k-1,k,1}\nonumber\\
&&+\tilde{L}_{5}^{k-1,k,1}\tilde{L}_{2}^{k-1,k,1}-
\tilde{L}_{4}^{k-1,k,1}\tilde{L}_{2}^{k-1,k,1}+
\tilde{L}_{3}^{k-1,k,1}\tilde{L}_{3}^{k-1,k,1})
\nonumber\\
&&+8\tilde{L}_{2}^{k-1,k,1}\tilde{L}_{3}^{k-1,k,2}
(\tilde{L}_{6}^{k-1,k,1}+\tilde{L}_{5}^{k-1,k,1}
-\tilde{L}_{3}^{k-1,k,1}-\tilde{L}_{2}^{k-1,k,1})\nonumber\\
&&-\frac{32}{3}(\tilde{L}_{2}^{k-1,k,1})^{3}(\tilde{L}_{6}^{k-1,k,1}+
2\tilde{L}_{5}^{k-1,k,1}
-2\tilde{L}_{3}^{k-1,k,1}-\tilde{L}_{2}^{k-1,k,1}).
\label{ex3}
\end{eqnarray}
Looking at these formulas, 
we assume three more characteristics of
$G_{d-m}^{N,k,d}(n;\sigma_{m})$.\\
(iv) 
\begin{equation}
G_{d}^{N,k,d}(n;(0))=\tilde{L}_{n}^{N,k,d}-\tilde{L}_{1+d(k-N)}^{N,k,d}.
\label{ass1}
\end{equation}
(v)
\begin{equation}
G_{d-m}^{N,k,d}(2+(k-N)(d+f);\sigma_{m})=
G_{d-m}^{N,k,d+f}(2+(k-N)(d+f);\sigma_{m}\cup(f)).
\label{ass2}
\end{equation}
(vi)
\begin{eqnarray}
&&G_{d-m}^{N,k,d}(n;(m))=\sum_{j=0}^{(k-N)(m)}
(\tilde{L}_{n-j}^{N,k,d-m}-\tilde{L}_{1+d(k-N)-j}^{N,k,d-m})\no\\
&&+(\mbox{homogeneous polynomials of degree $d-m$ that consists of 
$\tilde{L}_{l}^{N,k,m'}\;\;(m'< d-m )$}).\no\\
\label{ass3}
\end{eqnarray}
In view of the resolution diagram of ${\cal M}_{d}^{CP^{N-1}}$
in section 2, $G_{d}^{N,k,d}(n;(0))$
corresponds to the pair $CP^{N(d+1)-1}\rightarrow CP^{d}\times
CP^{N-1}$, that is the element with the largest dimension in the
diagram.
The condition (v) suggests that $G_{d-m}^{N,k,d}(n;\sigma_{m})$ inherits 
some combinatorial characteristics of the partitions of integers.   
  
\subsection{Solving the Ansatz (v) and 
Explicit Formula for the $d=3$ case}
In the first part of this section, we give a recipe (though it's 
not complete) of the inductive 
determination of the factor $G_{d-m}^{N,k,d}(n;\sigma_{m})$. 
We begin with the definition,
\begin{defi}
\begin{equation}
\tilde{G}_{d-m}^{N,k,d+f}(n;\sigma_{m}\cup(f)):=
\sum_{j=0}^{(k-N)f}G_{d-m}^{N,k,d}(n-j;\sigma_{m})-
\sum_{j=0}^{(k-N)f}G_{d-m}^{N,k,d}(1+(k-N)(d+f)-j;\sigma_{m}).
\label{gnew}
\end{equation}
\end{defi}
Then we are led to the following proposition,
\begin{prop}
$\tilde{G}_{d-m}^{N,k,d+f}(n;\sigma_{m}\cup(f))$
satisfy the ansatz (v):
\begin{equation}
\tilde{G}_{d-m}^{N,k,d+f}(2+(k-N)(d+f);\sigma_{m}\cup(f))
=G_{d-m}^{N,k,d}(2+(k-N)(d+f);\sigma_{m}),
\label{sol}
\end{equation}
and the ansatz (i), (ii), and (iii).
\end{prop}
{\it proof)} 
The fact that $\tilde{G}_{d-m}^{N,k,d+f}(n;\sigma_{m}\cup(f))$
satisfy the ansatz (i), (iii) is rather obvious,
and we first prove the ansatz (ii).
It suffices to consider the part
$\sum_{j=0}^{(k-N)f}G_{d-m}^{N,k,d}(n-j;\sigma_{m})$:
\begin{eqnarray}
&&\sum_{j=0}^{(k-N)f}G_{d-m}^{N,k,d}(N-1+(k-N)(d+f)-n-j;\sigma_{m})\no\\
&=&\sum_{j=0}^{(k-N)f}G_{d-m}^{N,k,d}(N-1+(k-N)d-n+j;\sigma_{m})\no\\
&=&\sum_{j=0}^{(k-N)f}G_{d-m}^{N,k,d}(n-j;\sigma_{m}).
\label{(ii)}
\end{eqnarray}  
Next, we turn to the formula 
$(\ref{sol})$. By definition and the condition
$G_{d-m}^{N,k,d}(1+(k-N)d;\sigma_{m})=0$, we obtain, 
\begin{eqnarray}
&&\tilde{G}_{d-m}^{N,k,d+f}(2+(k-N)(d+f);\sigma_{m}\cup(f))\no\\
&=&\sum_{j=0}^{(k-N)f}G_{d-m}^{N,k,d}(2+(k-N)(d+f)-j;\sigma_{m})
-\sum_{j=0}^{(k-N)f}G_{d-m}^{N,k,d}(1+(k-N)(d+f)-j;\sigma_{m})\no\\
&=&G_{d-m}^{N,k,d}(2+(k-N)(d+f);\sigma_{m}).
\label{prf}
\end{eqnarray}
 Q.E.D.
\begin{defi}
\begin{equation}
C_{d-m}^{N,k,d+f}(n;\sigma_{m}\cup(f)):=
G_{d-m}^{N,k,d+f}(n;\sigma_{m}\cup(f))-
\tilde{G}_{d-m}^{N,k,d+f}(n;\sigma_{m}\cup(f))
\label{difference}
\end{equation}
\end{defi}
\begin{prop}
$C_{d-m}^{N,k,d+f}(n;\sigma_{m}\cup(f))$ satisfies 
the condition:
\begin{equation}
C_{d-m}^{N,k,d+f}(2+(k-N)(d+f);\sigma_{m}\cup(f))
=C_{d-m}^{N,k,d+f}(1+(k-N)(d+f);\sigma_{m}\cup(f))=0,
\label{hidden}
\end{equation}
and the ansatz (i), (ii), (iii).
\end{prop}
{\it proof)} Immediate. Q.E.D.
\\

We can easily see that $C_{d-m}^{N,k,d+f}(n;\sigma_{m}\cup(f))$
consists of monomials of degree $d-m$ only in 
$\tilde{L}_{j}^{N,k,m'}\;\;(m'<d-m)$, and we are led to the corollary:
\begin{cor}
\begin{eqnarray}
G_{d-m}^{N,k,d}(n;\sigma_{m})&=&\sum_{j_{1}=0}^{d_{1}(k-N)}
\sum_{j_{2}=0}^{d_{2}(k-N)}\cdots
\sum_{j_{l(\sigma_{m})}=0}^{d_{l(\sigma_{m})}(k-N)}
\tilde{L}^{N,k,d-m}_{n-\sum_{i=1}^{l(\sigma_{m})}j_{i}}\no\\
&&-\sum_{j_{1}=0}^{d_{1}(k-N)}
\sum_{j_{2}=0}^{d_{2}(k-N)}\cdots
\sum_{j_{l(\sigma_{m})}=0}^{d_{l(\sigma_{m})}(k-N)}
\tilde{L}^{N,k,d-m}_{1+(k-N)d-\sum_{i=1}^{l(\sigma_{m})}j_{i}}\no\\
&&+(\mbox{ homogeneous polynomials of degree $d-m$ that consist of 
$\tilde{L}_{j}^{N,k,m'}\;\;(m'<d-m)$}).\no\\
\label{genetop1}
\end{eqnarray}
\end{cor} 
$C_{1}^{N,k,d}(n;\sigma_{d-f-1}\cup(f))$ must be zero just because 
there are no positive integers less than $1$. 
Hence we obtain, 
\begin{cor}
\begin{eqnarray}
G_{1}^{N,k,d}(n;\sigma_{d-1})&=&\sum_{j_{1}=0}^{d_{1}(k-N)}
\sum_{j_{2}=0}^{d_{2}(k-N)}\cdots
\sum_{j_{l(\sigma_{d-1})}=0}^{d_{l(\sigma_{d-1})}(k-N)}
\tilde{L}^{N,k,1}_{n-\sum_{i=1}^{l(\sigma_{d-1})}j_{i}}\no\\
&&-\sum_{j_{1}=0}^{d_{1}(k-N)}
\sum_{j_{2}=0}^{d_{2}(k-N)}\cdots
\sum_{j_{l(\sigma_{d-1})}=0}^{d_{l(\sigma_{d-1})}(k-N)}
\tilde{L}^{N,k,1}_{1+(k-N)d-\sum_{i=1}^{l(\sigma_{d-1})}j_{i}}.
\label{genetop2}
\end{eqnarray}
\end{cor} 
Note that these formulas indeed imply a close relation between 
$G_{d-m}^{N,k,d}(n;\sigma_{m})$ and the partition 
$\sigma_{m}$. After all, what remains to be determined is the ``hidden 
intersections'' $C_{d-m}^{N,k,d+f}(n;\sigma_{m}\cup(f))$ that appear 
in the inductive process. We would like to give the general treatment 
of them in the future work, but in the $d=3$ case, we managed to
determine them using the condition (\ref{hidden}) and the numerical
results as will be discussed in the latter half of this section.
    
Now, we apply the disscussion so far to the case of $d=3$.
In this case, we have to determine the following four 
factors:
\begin{eqnarray}
&&G_{3}^{N,k,3}(n;(0)),\;\;\;G_{2}^{N,k,3}(n;(1)) \no\\
&&G_{1}^{N,k,3}(n;(1)+(1)),\;\;\;G_{1}^{N,k,3}(n;(2))
\end{eqnarray}

Straightforward application of the results so far leads us 
to the following conjecture.
\begin{conj}
 The mirror transformation for  $d=3$ is given by
\begin{eqnarray}
L_{n}^{N,k,3}&=&\tilde{L}_{n}^{N,k,3}-\tilde{L}_{1+3(k-N)}^{N,k,3}
-3\tilde{L}_{1+(k-N)}^{N,k,1}(\sum_{j=0}^{k-N}
(\tilde{L}_{n-j}^{N,k,2}-\tilde{L}_{1+3(k-N)-j}^{N,k,2})
+C_{1,1}^{N,k,3}(n))\nonumber\\
&&-\frac{3}{2}\tilde{L}_{1+2(k-N)}^{N,k,2}
(\sum_{j=0}^{2(k-N)}(\tilde{L}_{n-j}^{N,k,1}
-\tilde{L}_{1+3(k-N)-j}^{N,k,1}))\no\\
&&+\frac{9}{2}(\tilde{L}_{1+(k-N)}^{N,k,1})^{2}
(\sum_{j=0}^{2(k-N)}A_{j}(\tilde{L}_{n-j}^{N,k,1}
-\tilde{L}_{1+3(k-N)-j}^{N,k,1})),
\label{cubic}
\end{eqnarray}
where
\begin{equation}
A_{j}:=j+1,\;\;\mbox{if}\;\;\;(0\leq j\leq k-N),\;\;\;
A_{j}:=1+2(k-N)-j,\;\;\mbox{if}\;\;\; (k-N\leq j\leq 2(k-N)).
\end{equation}
and $C_{1,1}^{N,k,3}(n)$ is some degree $2$ polynomial of 
$\tilde{L}_{n-j}^{N,k,1},\;(0\leq j \leq
2(k-N))$, and of $\tilde{L}_{1+3(k-N)-j}^{N,k,1},\;(0\leq j \leq 2(k-N))$,
that satisfy
\begin{eqnarray}
&&C_{1,1}^{N,k,3}(n)=C_{1,1}^{N,k,3}(N-1+3(k-N)-n),\no\\
&&C_{1,1}^{N,k,3}(1+3(k-N))=C_{1,1}^{N,k,3}(2+3(k-N))=0.
\end{eqnarray}
\end{conj}
As we mentioned before,  
the hidden intersection $C_{1,1}^{N,k,3}(n)$ is yet to be determined. 
For simplicity, we discuss the case of $k-N=1$ in detail.
 
By the flat metric condition and 
the condition (iii), $C_{1,1}^{k-1,k,3}(n)$ must have the form
\begin{eqnarray}
&&\sum_{i=0}^{2}\sum_{j=0}^{2}A_{ij}\tilde{L}_{n-i}^{k-1,k,1}
\tilde{L}_{n-j}^{k-1,k,1}+\sum_{i=0}^{2} 
\sum_{j=0}^{2}B_{ij}\tilde{L}_{4-i}^{k-1,k,1}
\tilde{L}_{n-j}^{k-1,k,1}\no\\
&&-(\sum_{i=0}^{2}\sum_{j=0}^{2}A_{ij}\tilde{L}_{4-i}^{k-1,k,1}
\tilde{L}_{4-j}^{k-1,k,1}+\sum_{i=0}^{2} 
\sum_{j=0}^{2}B_{ij}\tilde{L}_{4-i}^{k-1,k,1}
\tilde{L}_{4-j}^{k-1,k,1}).
\label{c11}
\end{eqnarray}
We further impose the symmetry condition and the condition 
$C_{1,1}^{k-1,k,3}(5)=0$. Then we obtain
\begin{eqnarray}
&&C_{1,1}^{k-1,k,3}(n)=C_{1}(\tilde{L}_{n}^{k-1,k,1}\tilde{L}_{n-2}^{k-1,k,1}-
\tilde{L}_{3}^{k-1,k,1}
(\tilde{L}_{n}^{k-1,k,1}+\tilde{L}_{n-1}^{k-1,k,1}+
\tilde{L}_{n-2}^{k-1,k,1})+
\tilde{L}_{2}^{k-1,k,1}(\tilde{L}_{n-1}^{k-1,k,1})\no\\
&&-(\tilde{L}_{4}^{k-1,k,1}\tilde{L}_{2}^{k-1,k,1}-
\tilde{L}_{3}^{k-1,k,1}
(\tilde{L}_{4}^{k-1,k,1}+\tilde{L}_{3}^{k-1,k,1}+
\tilde{L}_{2}^{k-1,k,1})
+\tilde{L}_{2}^{k-1,k,1}(\tilde{L}_{3}^{k-1,k,1})))\no\\
&&+C_{2}(\tilde{L}_{n-1}^{k-1,k,1}\tilde{L}_{n-1}^{k-1,k,1}-
(\tilde{L}_{4}^{k-1,k,1}+\tilde{L}_{3}^{k-1,k,1})\tilde{L}_{n-1}^{k-1,k,1}
\no\\
&&-(\tilde{L}_{3}^{k-1,k,1}\tilde{L}_{3}^{k-1,k,1}-
(\tilde{L}_{4}^{k-1,k,1}+\tilde{L}_{3}^{k-1,k,1})\tilde{L}_{3}^{k-1,k,1})).
\label{c1c2}
\end{eqnarray}
If we compare (\ref{c1c2}) with (\ref{ex2}), we can see 
$C_{1}=1, C_{2}=0$.
Similarly, we determined $C_{1,1}^{N,k,3}(n)$ for the $k-N=1,2$ case 
using numerical results for lower $N$ and $k$.
\begin{eqnarray}
C_{1,1}^{k-1,k,3}(n)&=&\tilde{L}_{n}^{k-1,k,1}\tilde{L}_{n-2}^{k-1,k,1}-
\tilde{L}_{3}^{k-1,k,1}
(\tilde{L}_{n}^{k-1,k,1}+\tilde{L}_{n-1}^{k-1,k,1}+
\tilde{L}_{n-2}^{k-1,k,1})
+\tilde{L}_{2}^{k-1,k,1}(\tilde{L}_{n-1}^{k-1,k,1})\no\\
&&-\tilde{L}_{4}^{k-1,k,1}\tilde{L}_{2}^{k-1,k,1}+
\tilde{L}_{3}^{k-1,k,1}
(\tilde{L}_{4}^{k-1,k,1}+\tilde{L}_{3}^{k-1,k,1}+
\tilde{L}_{2}^{k-1,k,1})
-\tilde{L}_{2}^{k-1,k,1}(\tilde{L}_{3}^{k-1,k,1})\no\\
C_{1,1}^{k-2,k,3}(n)&=&\tilde{L}_{n}^{k-2,k,1}\tilde{L}_{n-4}^{k-2,k,1}
-\tilde{L}_{4}^{k-2,k,1}
(\tilde{L}_{n}^{k-2,k,1}+\tilde{L}_{n-1}^{k-2,k,1}+
\tilde{L}_{n-2}^{k-2,k,1}+\tilde{L}_{n-3}^{k-2,k,1}+
\tilde{L}_{n-4}^{k-2,k,1})\no\\
&&+\tilde{L}_{3}^{k-2,k,1}(\tilde{L}_{n-1}^{k-2,k,1}
+\tilde{L}_{n-2}^{k-2,k,1}+\tilde{L}_{n-3}^{k-2,k,1})\no\\
&&-\tilde{L}_{7}^{k-2,k,1}\tilde{L}_{3}^{k-2,k,1}
+\tilde{L}_{4}^{k-2,k,1}
(\tilde{L}_{7}^{k-2,k,1}+\tilde{L}_{6}^{k-2,k,1}+
\tilde{L}_{5}^{k-2,k,1}+\tilde{L}_{4}^{k-2,k,1}+
\tilde{L}_{3}^{k-2,k,1})\no\\
&&-\tilde{L}_{3}^{k-2,k,1}(\tilde{L}_{6}^{k-2,k,1}
+\tilde{L}_{5}^{k-2,k,1}+\tilde{L}_{4}^{k-2,k,1})\no\\
&&+\tilde{L}_{n}^{k-2,k,1}\tilde{L}_{n-3}^{k-2,k,1}
+\tilde{L}_{n-1}^{k-2,k,1}\tilde{L}_{n-4}^{k-2,k,1}\no\\
&&-\tilde{L}_{5}^{k-2,k,1}
(\tilde{L}_{n}^{k-2,k,1}+\tilde{L}_{n-1}^{k-2,k,1}+
\tilde{L}_{n-2}^{k-2,k,1}+\tilde{L}_{n-3}^{k-2,k,1}+
\tilde{L}_{n-4}^{k-2,k,1})
+\tilde{L}_{3}^{k-2,k,1}(\tilde{L}_{n-2}^{k-2,k,1})\no\\
&&-\tilde{L}_{7}^{k-2,k,1}\tilde{L}_{4}^{k-2,k,1}
-\tilde{L}_{6}^{k-2,k,1}\tilde{L}_{3}^{k-2,k,1}\no\\
&&+\tilde{L}_{5}^{k-2,k,1}
(\tilde{L}_{7}^{k-2,k,1}+\tilde{L}_{6}^{k-2,k,1}+
\tilde{L}_{5}^{k-2,k,1}+\tilde{L}_{4}^{k-2,k,1}+
\tilde{L}_{3}^{k-2,k,1})
-\tilde{L}_{3}^{k-2,k,1}(\tilde{L}_{5}^{k-2,k,1})\no\\
\label{quad} 
\end{eqnarray}
From (\ref{quad}), we can speculate the form of $C_{1,1}^{N,k,3}(n)$ 
as
\begin{eqnarray}
C_{1,1}^{N,k,3}(n)&=&\sum_{j=0}^{(k-N)-1}(\sum_{m=0}^{j}\tilde{L}_{n-m}^{N,k,1}
\tilde{L}_{n-2(k-N)+j-m}^{N,k,1}-\tilde{L}_{(k-N)+2+j}^{N,k,1}
(\sum_{m=0}^{2(k-N)}\tilde{L}_{n-m}^{N,k,1})\no\\
&&+\tilde{L}_{1+(k-N)}^{N,k,1}
(\sum_{m=j+1}^{2(k-N)-j-1}\tilde{L}_{n-m}^{N,k,1}))\no\\
&&-\sum_{j=0}^{(k-N)-1}(\sum_{m=0}^{j}\tilde{L}_{1+3(k-N)-m}^{N,k,1}
\tilde{L}_{1+(k-N)+j-m}^{N,k,1}-\tilde{L}_{(k-N)+2+j}^{N,k,1}
(\sum_{m=0}^{2(k-N)}\tilde{L}_{1+3(k-N)-m}^{N,k,1})\no\\
&&+\tilde{L}_{1+(k-N)}^{N,k,1}
(\sum_{m=j+1}^{2(k-N)-j-1}\tilde{L}_{1+3(k-N)-m}^{N,k,1})).
\label{fini}
\end{eqnarray}
We want to show that (\ref{cubic}) and (\ref{fini}) correctly predicts 
$L_{n}^{N,k,3}$ for all $N$ and $k$. We did numerical check for
some lower $N$ and $k$ by the method of toroidal 
calculation considered by Kontsevich \cite{tor} and no contradiction occurred. 
\subsection{Partial Results for the $d=4$ case}
In the $d=4$ case, we can partially 
determine the form of the generalized mirror transformation, 
using our results so far, as follows: 
\begin{eqnarray}
L_{n}^{N,k,4}&=&\tilde{L}_{n}^{N,k,4}-\tilde{L}_{1+4(k-N)}^{N,k,4}
\nonumber\\
&&-4\tilde{L}_{1+(k-N)}^{N,k,1}(\sum_{j=0}^{(k-N)}(\tilde{L}_{n-j}^{N,k,3}
-\tilde{L}_{1+4(k-N)-j}^{N,k,3})
+C^{N,k,4}_{2,1}(n)+C^{N,k,4}_{1,1,1}(n))\nonumber\\
&&-2\tilde{L}_{1+2(k-N)}^{N,k,2}(\sum_{j=0}^{2(k-N)}(\tilde{L}_{n-j}^{N,k,2}
-\tilde{L}_{1+4(k-N)-j}^{N,k,2})+C^{N,k,4}_{1,1}(n))\nonumber\\
&&-\frac{4}{3}\tilde{L}_{1+3(k-N)}^{N,k,3}(\sum_{j=0}^{3(k-N)}( 
\tilde{L}_{n-j}^{N,k,1}-\tilde{L}_{1+4(k-N)-j}^{N,k,1}))
\nonumber\\
&&+8(\tilde{L}_{1+(k-N)}^{N,k,1})^{2}(\sum_{j=0}^{2(k-N)}
A_{j}(\tilde{L}_{n-j}^{N,k,2}
-\tilde{L}_{1+4(k-N)-j}^{N,k,2})\no\\
&&+\sum_{j=0}^{k-N}(E^{N,k,3}_{1,1}(n-j)-E^{N,k,3}_{1,1}(1+4(k-N)-j))
+D^{N,k,4}_{1,1}(n))
\nonumber\\
&&+8\tilde{L}_{1+(k-N)}^{N,k,1}\tilde{L}_{1+2(k-N)}^{N,k,2}
(\sum_{j=0}^{3(k-N)}B_{j}(\tilde{L}_{n-j}^{N,k,1}-
\tilde{L}_{1+4(k-N)-j}^{N,k,1}))\nonumber\\
&&-\frac{32}{3}(\tilde{L}_{1+(k-N)}^{N,k,1})^{3}
(\sum_{j=0}^{3(k-N)}C_{j}(\tilde{L}_{n-j}^{N,k,1}-
\tilde{L}_{1+4(k-N)-j}^{N,k,1})),
\label{4}
\end{eqnarray}
where $A_{j}$ is the same as defined in Conjecture 2 and
\begin{eqnarray}
B_{j}&=&j+1\;\;\;\;\;(0\leq j\leq k-N),\no\\
&=&k-N+1\;\;\;\;(k-N\leq j\leq 2(k-N)),\no\\
&=&3(k-N)+1-j\;\;\;\;(2(k-N)\leq j\leq 3(k-N)),\no\\
C_{j}&=&\frac{(j+1)(j+2)}{2}\;\;\;\;(0\leq j\leq k-N),\no\\
&=&-j^{2}+3(k-N)j-\frac{3}{2}(k-N)(k-N-1)+1\;\;\;(k-N\leq j\leq 2(k-N)),\no\\
&=&\frac{(3(k-N)-j+1)(3(k-N)-j+2)}{2}\;\;\;(2(k-N)\leq j\leq 3(k-N)),\no\\
E^{N,k,3}_{1,1}(n)&=&\sum_{j=0}^{(k-N)-1}(\sum_{m=0}^{j}\tilde{L}_{n-m}^{N,k,1}
\tilde{L}_{n-2(k-N)+j-m}^{N,k,1}-\tilde{L}_{(k-N)+2+j}^{N,k,1}
(\sum_{m=0}^{2(k-N)}\tilde{L}_{n-m}^{N,k,1})\no\\
&&+\tilde{L}_{1+(k-N)}^{N,k,1}
(\sum_{m=j+1}^{2(k-N)-j-1}\tilde{L}_{n-m}^{N,k,1})).
\end{eqnarray}
$C_{2,1}^{N,k,4}(n)$, $C_{1,1,1}^{N,k,4}(n)$, 
$C_{1,1}^{N,k,4}(n)$ and $D_{1,1}^{N,k,4}(n)$ are ``hidden
intersections'' and are some weighted homogeneous 
polynomials that respectively consist of 
$\tilde{L}_{i}^{N,k,2}\tilde{L}_{j}^{N,k,1}$, 
$\tilde{L}_{i}^{N,k,1}\tilde{L}_{j}^{N,k,1}\tilde{L}_{l}^{N,k,1}$,
$\tilde{L}_{i}^{N,k,1}\tilde{L}_{j}^{N,k,1}$ and 
$\tilde{L}_{i}^{N,k,1}\tilde{L}_{j}^{N,k,1}$, 
which satisfy
\begin{eqnarray}
&&C_{2,1}^{N,k,4}(1+4(k-N))=C_{1,1,1}^{N,k,4}(1+4(k-N))
=C_{1,1}^{N,k,4}(1+4(k-N))\no\\
&&=C_{2,1}^{N,k,4}(2+4(k-N))=C_{1,1,1}^{N,k,4}(2+4(k-N))
=C_{1,1}^{N,k,4}(2+4(k-N))\no\\
&&=D_{1,1}^{N,k,4}(1+4(k-N))=D_{1,1}^{N,k,4}(2+4(k-N))
=0,\no\\
&&C_{2,1}^{N,k,4}(n)=C_{2,1}^{N,k,4}(N-1+4(k-N)-n),\no\\
&&C_{1,1,1}^{N,k,4}(n)=C_{1,1,1}^{N,k,4}(N-1+4(k-N)-n),\no\\
&&C_{1,1}^{N,k,4}(n)=C_{1,1}^{N,k,4}(N-1+4(k-N)-n),\no\\
&&D_{1,1}^{N,k,4}(n)=D_{1,1}^{N,k,4}(N-1+4(k-N)-n). 
\end{eqnarray}
If we set $n=2+4(k-N)$, the formula (\ref{4}) includes no ambiguous 
part and we obtain the following proposition.   
\begin{prop}
In the case of $n=2+4(k-N)$, the following formula predicts 
$L_{2+4(k-N)}^{N,k,4}$ for arbitrary $N$ and $k$.
\begin{eqnarray}
L_{2+4(k-N)}^{N,k,4}
&=&\tilde{L}_{2+4(k-N)}^{N,k,4}-\tilde{L}_{1+4(k-N)}^{N,k,4}
\nonumber\\
&&-4\tilde{L}_{1+(k-N)}^{N,k,1}(\tilde{L}_{2+4(k-N)}^{N,k,3}
-\tilde{L}_{1+3(k-N)}^{N,k,3})\nonumber\\
&&-2\tilde{L}_{1+2(k-N)}^{N,k,2}(\tilde{L}_{2+4(k-N)}^{N,k,2}
-\tilde{L}_{1+2(k-N)}^{N,k,2})\nonumber\\
&&-\frac{4}{3}\tilde{L}_{1+3(k-N)}^{N,k,3}( 
\tilde{L}_{2+4(k-N)}^{N,k,1}-\tilde{L}_{1+(k-N)}^{N,k,1})
\nonumber\\
&&+8(\tilde{L}_{1+(k-N)}^{N,k,1})^{2}(\sum_{j=0}^{(k-N)}
(\tilde{L}_{2+4(k-N)-j}^{N,k,2}
-\tilde{L}_{1+3(k-N)-j}^{N,k,2})+C^{N,k,3}_{1,1}(2+4(k-N)))
\nonumber\\
&&+8\tilde{L}_{1+(k-N)}^{N,k,1}\tilde{L}_{1+2(k-N)}^{N,k,2}
(\sum_{j=0}^{2(k-N)}(\tilde{L}_{2+4(k-N)-j}^{N,k,1}-
\tilde{L}_{1+3(k-N)-j}^{N,k,1}))\nonumber\\
&&-\frac{32}{3}(\tilde{L}_{1+(k-N)}^{N,k,1})^{3}
(\sum_{j=0}^{2(k-N)}A_{j}(\tilde{L}_{2+4(k-N)-j}^{N,k,1}-
\tilde{L}_{1+3(k-N)-j}^{N,k,1})).
\end{eqnarray}
\end{prop}
{\bf Remark 1}
From numerical computations for lower $N$ and $k$,
we determined $C_{1,1}^{N,k,4}(n)$ using the fact that 
$-4\tilde{L}_{1+(k-N)}^{N,k,1}(C^{N,k,4}_{2,1}(n)+C^{N,k,4}_{1,1,1}(n))+
8(\tilde{L}_{1+(k-N)}^{N,k,1})^{2}D^{N,k,4}_{1,1}(n)$ is 
divisible by $\tilde{L}_{1+(k-N)}^{N,k,1}$.
\begin{eqnarray}
C_{1,1}^{N,k,4}(n)&=&\sum_{j=0}^{2(k-N)-1}
(\sum_{m=0}^{j}\tilde{L}_{n-m}^{N,k,1}
\tilde{L}_{n-3(k-N)+j-m}^{N,k,1}-\tilde{L}_{(k-N)+2+j}^{N,k,1}
(\sum_{m=0}^{3(k-N)}\tilde{L}_{n-m}^{N,k,1}))\no\\
&&-\sum_{j=0}^{2(k-N)-1}
(\sum_{m=0}^{j}\tilde{L}_{1+4(k-N)-m}^{N,k,1}
\tilde{L}_{1+(k-N)+j-m}^{N,k,1}-\tilde{L}_{(k-N)+2+j}^{N,k,1}
(\sum_{m=0}^{3(k-N)}\tilde{L}_{1+4(k-N)-m}^{N,k,1}))\no\\
&&+\tilde{L}_{1+(k-N)}^{N,k,1}
(\sum_{j=0}^{k-N-1}\sum_{m=j+1}^{3(k-N)-j-1}(\tilde{L}_{n-m}^{N,k,1}
-\tilde{L}_{1+4(k-N)-m}^{N,k,1})).
\label{fini2}
\end{eqnarray}
This formula satisfies the condition 
$C_{1,1}^{N,k,4}(2+5(k-N))=C_{1,1}^{N,k,3}(2+5(k-N))$, that is 
imposed by the ansatz (v).
\section{Some Examples}
In this section, we show some explicit examples of the quantum cohomology of 
general type hypersurfaces and the generalized mirror transformation.
We put the real structure constants on the l.h.s.of the generalized 
mirror transformation.
On the r.h.s. of the equalities, we put the virtual 
structure constants obtained by applying the recursive formulas 
in Appendix B with the initial conditions listed in Table 1.   
Some other examples of the generalized mirror transformation 
are listed in Table 2 and Table 3.\\
$M_{6}^{7}$ model\\
{\bf Multiplication Rule}
\begin{eqnarray}
{\cal O}_{e}\cdot 1 &=& {\cal O}_{e},\nonumber\\
{\cal O}_{e}\cdot{\cal O}_{e}&=&{\cal O}_{{e}^2}+
A{\cal O}_{{e}^3}q,
 \nonumber\\
{\cal O}_{e}\cdot{\cal O}_{{e}^2}&=&{\cal O}_{{e}^3},
 \nonumber\\
{\cal O}_{e}\cdot{\cal O}_{{e}^3}&=&{\cal O}_{{e}^4},
 \nonumber\\
{\cal O}_{e}\cdot{\cal O}_{{e}^4}&=&0.
\label{mul}
\end{eqnarray}
where $A=99715$\\
{\bf Generalized Mirror Transformation}
\begin{equation}
99715=300167-200452
\end{equation}
{\bf Generator Representation}
\begin{eqnarray}
{\cal O}_{e} &=& {\cal O}_{e},\nonumber\\
{\cal O}_{e^{2}}&=&({\cal O}_{e})^{2}\over
{1+A{\cal O}_{e}q},
 \nonumber\\
{\cal O}_{e^{3}}&=&({\cal O}_{e})^{3}\over
{1+A{\cal O}_{e}q},
 \nonumber\\
{\cal O}_{e^{4}}&=&({\cal O}_{e})^{4}\over
{1+A{\cal O}_{e}q}.
\label{gen}
\end{eqnarray}
{\bf Relation}
\begin{equation}
\frac{({\cal O}_{e})^{5}}{(1+A{\cal O}_{e}q)}=0.
\label{rel}
\end{equation}
$M_{7}^{8}$ model\\
{\bf Multiplication Rule}
\begin{eqnarray}
{\cal O}_{e}\cdot 1 &=& {\cal O}_{e},\nonumber\\
{\cal O}_{e}\cdot{\cal O}_{e}&=&{\cal O}_{{e}^2}+
A{\cal O}_{{e}^3}q+B{\cal O}_{{e}^4}q^{2},
 \nonumber\\
{\cal O}_{e}\cdot{\cal O}_{{e}^2}&=&{\cal O}_{{e}^3}
+A{\cal O}_{{e}^4}q,\nonumber\\
{\cal O}_{e}\cdot{\cal O}_{{e}^3}&=&{\cal O}_{{e}^4},
 \nonumber\\
{\cal O}_{e}\cdot{\cal O}_{{e}^4}&=&{\cal O}_{{e}^5},
\nonumber\\
{\cal O}_{e}\cdot{\cal O}_{{e}^5}&=&0.
\label{mul2}
\end{eqnarray}
where $A=2689792,  B=21553860841856$.\\
{\bf Generalized Mirror Transformation}
\begin{eqnarray}
2689792&=&5241984-2552192,\nonumber\\
21553860841856&=&91173486748800-55889894658816\nonumber\\
&&-2*2552192*(5241984+5241984-5241984-2552192). 
\end{eqnarray}
{\bf Generator Representation}
\begin{eqnarray}
{\cal O}_{e} &=& {\cal O}_{e},\nonumber\\
{\cal O}_{e^{2}}&=&{(1+A{\cal O}_{e}q)({\cal O}_{e})^{2}}\over
{1+2A{\cal O}_{e}q+B({\cal O}_{e})^{2}q^{2}},
 \nonumber\\
{\cal O}_{e^{3}}&=&({\cal O}_{e})^{3}\over
{1+2A{\cal O}_{e}q+B({\cal O}_{e})^{2}q^{2}},
 \nonumber\\
{\cal O}_{e^{4}}&=&({\cal O}_{e})^{4}\over
{1+2A{\cal O}_{e}q+B({\cal O}_{e})^{2}q^{2}},
 \nonumber\\
{\cal O}_{e^{5}}&=&({\cal O}_{e})^{5}\over
{1+2A{\cal O}_{e}q+B({\cal O}_{e})^{2}q^{2}}.
\label{gen2}
\end{eqnarray}
{\bf Relation}
\begin{equation}
\frac{({\cal O}_{e})^{6}}{1+2A{\cal O}_{e}q+B({\cal O}_{e})^{2}q^{2}}=0.
\label{rel2}
\end{equation}
$M_{8}^{9}$ model\\
{\bf Multiplication Rule}
\begin{eqnarray}
{\cal O}_{e}\cdot 1 &=& {\cal O}_{e},\nonumber\\
{\cal O}_{e}\cdot{\cal O}_{e}&=&{\cal O}_{{e}^2}+
A{\cal O}_{{e}^3}q+B{\cal O}_{{e}^4}q^{2}+C{\cal O}_{{e}^5}q^{3},
 \nonumber\\
{\cal O}_{e}\cdot{\cal O}_{{e}^2}&=&{\cal O}_{{e}^3}
+D{\cal O}_{{e}^4}q+B{\cal O}_{{e}^5}q^{2},\nonumber\\
{\cal O}_{e}\cdot{\cal O}_{{e}^3}&=&{\cal O}_{{e}^4}
+A{\cal O}_{{e}^5}q,\nonumber\\
{\cal O}_{e}\cdot{\cal O}_{{e}^4}&=&{\cal O}_{{e}^5},
\nonumber\\
{\cal O}_{e}\cdot{\cal O}_{{e}^5}&=&{\cal O}_{{e}^6},
\nonumber\\
{\cal O}_{e}\cdot{\cal O}_{{e}^6}&=&0.  
\label{mul3}
\end{eqnarray}
where $A=56718144, B=35512880615374365/2, C=4037555975532386945225553,
D=90617373$.\\
{\bf Generalized Mirror Transformation}
\begin{eqnarray}
56718144&=&90857052-34138908,\nonumber\\
90617373&=&124756281-34138908,\nonumber\\
35512880615374365/2&=&81506931029963973/2-16809868887197436\nonumber\\
&&-2*34138908*(124756281+90857052-90857052-34138908), \nonumber\\ 
4037555975532386945225553&=&
18892465499391490557425853-11447799161101387518646386\nonumber\\
&&-3*34138908*(81506931029963973/2-16809868887197436)\nonumber\\
&&-(3/2)*16809868887197436*(90857052-34138908)\no\\
&&+(9/2)*34138908*34138908*(124756281-34138908).
\end{eqnarray}
{\bf Generator Representation}
\begin{eqnarray}
{\cal O}_{e} &=& {\cal O}_{e},\nonumber\\
{\cal O}_{e^{2}}&=&{(1+(A+D){\cal O}_{e}q+B({\cal O}_{e})^{2}q^{2})
({\cal O}_{e})^{2}}\over
{1+(2A+D){\cal O}_{e}q+(2B+A^{2})({\cal O}_{e})^{2}q^{2}+
C({\cal O}_{e})^{3}q^{3}},
 \nonumber\\
{\cal O}_{e^{3}}&=&{(1+A{\cal O}_{e}q)({\cal O}_{e})^{3}}\over
{1+(2A+D){\cal O}_{e}q+(2B+A^{2})({\cal O}_{e})^{2}q^{2}+
C({\cal O}_{e})^{3}q^{3}},
 \nonumber\\
{\cal O}_{e^{4}}&=&({\cal O}_{e})^{4}\over
{1+(2A+D){\cal O}_{e}q+(2B+A^{2})({\cal O}_{e})^{2}q^{2}+
C({\cal O}_{e})^{3}q^{3}},
 \nonumber\\
{\cal O}_{e^{5}}&=&({\cal O}_{e})^{5}\over
{1+(2A+D){\cal O}_{e}q+(2B+A^{2})({\cal O}_{e})^{2}q^{2}+
C({\cal O}_{e})^{3}q^{3}},
\nonumber\\
{\cal O}_{e^{6}}&=&({\cal O}_{e})^{6}\over
{1+(2A+D){\cal O}_{e}q+(2B+A^{2})({\cal O}_{e})^{2}q^{2}+
C({\cal O}_{e})^{3}q^{3}}.
\label{gen3}
\end{eqnarray}
{\bf Relation}
\begin{equation}
\frac{({\cal O}_{e})^{7}}
{1+(2A+D){\cal O}_{e}q+(2B+A^{2})({\cal O}_{e})^{2}q^{2}+
C({\cal O}_{e})^{3}q^{3}}=0.
\label{rel3}
\end{equation}
\section{Conclusion}
In this paper, we have discussed a generalization of the mirror calculation of 
Calabi-Yau hypersurfaces to the case of general type hypersurfaces. 
However, our determination 
of the factor $G_{d-m}^{N,k,d}(n;\sigma_{m})$ relies 
on numerical results and depends on power of computers. 
That's why we stopped up to the partial results for the $d=4$ rational 
curves. So we have to search for other characteristics of 
$G_{d-m}^{N,k,d}(n;\sigma_{m})$ that is enough for determination of 
them. Geometrical derivation of our formula from the point of view 
of the projective space resolution in section 2 should be pursued further. 
Heuristically, we can expect that the pair 
\begin{equation}
CP^{d_{\pi(1)}}\times \cdots \times
CP^{d_{\pi(l(\sigma_{m}))}}\times CP^{N(d-m+1)-1}\rightarrow 
CP^{d_{\pi(1)}}\times\cdots \times
CP^{d_{\pi(l(\sigma_{m}))}}\times CP^{d-m}\times CP^{N-1},
\label{concl}
\end{equation}
$(\pi\in S_{l(\sigma_{m})})$,
in the resolution diagram corresponds to 
\begin{eqnarray}
&&\frac{d^{l(\sigma_{m})}}
{(\prod_{j=1}^{l(\sigma_{m})}d_{j})l(\sigma_{m})!}
\prod_{j=1}^{l(\sigma_{m})}\tilde{L}^{N,k,d_{j}}_{1+(k-N)d_{j}}
G_{d-m}^{N,k,d}(n;\sigma_{m})
\label{spond}
\end{eqnarray}
in the generalized mirror transformation.
Moreover it seems that $CP^{d_{j}}$ and the pair 
$CP^{N(d-m+1)-1}\rightarrow CP^{d-m}\times CP^{N-1}$ 
produce $\tilde{L}^{N,k,d_{j}}_{1+(k-N)d_{j}}$ and 
$G_{d-m}^{N,k,d}(n;\sigma_{m})$ respectively.
Geometrical derivation of the factor $\frac{d^{l(\sigma_{m})}}
{(\prod_{j=1}^{l(\sigma_{m})}d_{j})l(\sigma_{m})!}$ should  also 
be considered deeply.
These approaches will reveal the interaction between the geometrical 
structure of the moduli space of the holomorphic maps from $CP^{1}$
to the K\"ahler manifold of general type 
and the structure of the oscillator algebra of the free bosons 
(or the ring of the symmetric functions).
 We hope our approach will contribute to 
deeper understanding of the mirror phenomena, which is now generalized 
to various types of manifolds, not only in the case of Calabi-Yau 
manifolds.     

{\bf Acknowledgment}
We would like to thank A.Collino, S.Hosono, 
A.Matsuo, Y.Matsuo, T.Sugimoto  and
 T.Eguchi for discussions and kind encouragement.
The author is supported by grant of Japan Society for Promotion of 
Science. Numerical calculation of this paper is done by use of 
Mathematica.
\newpage
\section*{ Appendix A:Proof of Theorem 2}
In this section, we give the proof of Theorem 2. Here 
we keep the notation used in the proof of Proposition 1.
\begin{lem}
For any formal power series $f(z)=\sum_{j=1}^{\infty}u_{j}z^{j}$, 
the following formal power series identity holds. 
\begin{equation}
\sum_{d=1}^{\infty}\frac{1}{d}Res_{z=0}(z^{-d-1}
\exp(-df(z)))\cdot z^{d}\exp(df(z))=-f(z)
\end{equation}
\end{lem}
{\it proof)} First, let 
$S(\sigma_d)$ be defined by  
\begin{equation}
\exp(\sum_{j=1}^{\infty}u_{j}z^{j})=\sum_{d=0}^{\infty}
(\sum_{\sigma_d\in P_{d}}S(\sigma_d)
\prod_{j=1}^{l(\sigma_{d})}u_{d_{j}})z^{d}.
\label{not1}
\end{equation}
Then the above identity is equivalent to the combinatorial 
relations 
\begin{eqnarray}
\sum_{f=1}^{d}
\sum_{{\sigma_{f}\in P_{f},\sigma_{d-f}\in P_{d-f}}\atop{\sigma_{f}\cup
\sigma_{d-f}=\sigma_{d}}}(-1)^{l(\sigma_{f})}f^{l(\sigma_{d})-1}
S(\sigma_{f})S(\sigma_{d-f})&=&-1,\;\;(l(\sigma_{d})=1),\no\\
\sum_{f=1}^{d}
\sum_{{\sigma_{f}\in P_{f},\sigma_{d-f}\in P_{d-f}}\atop{\sigma_{f}\cup
\sigma_{d-f}=\sigma_{d}}}(-1)^{l(\sigma_{f})}f^{l(\sigma_{d})-1}
S(\sigma_{f})S(\sigma_{d-f})&=&0,\;\;(\mbox{otherwise}).
\label{comb1}
\end{eqnarray}
They are further unified into the relation,
\begin{eqnarray}
\sum_{f=0}^{d}
\sum_{{\sigma_{f}\in P_{f},\sigma_{d-f}\in P_{d-f}}\atop{\sigma_{f}\cup
\sigma_{d-f}=\sigma_{d}}}(-1)^{l(\sigma_{f})}f^{l(\sigma_{d})-1}
S(\sigma_{f})S(\sigma_{d-f})&=&0.
\label{comb2}
\end{eqnarray}
Hence it suffices to show
\begin{eqnarray}
\sum_{f=0}^{d}
\sum_{{\sigma_{f}\in P_{f},\sigma_{d-f}\in P_{d-f}}\atop{\sigma_{f}\cup
\sigma_{d-f}=\sigma_{d}}}(-1)^{l(\sigma_{f})}f^{m}
S(\sigma_{f})S(\sigma_{d-f})&=&0,\;\;\;(m\leq l(\sigma_{d})-1).
\label{comb3}
\end{eqnarray}
The l.h.s. of (\ref{comb3}) is the coefficient of 
$(\prod_{j=1}^{l(\sigma_{d})}u_{d_{j}})z^{d}$ in 
\begin{equation}
((z\frac{d}{dz})^{m}\exp(-\sum_{j=1}^{\infty}u_{j}z^{j}))\cdot
\exp(\sum_{j=1}^{\infty}u_{j}z^{j})).
\label{exp}
\end{equation}
But (\ref{exp}) obviously contains no monomial 
$(\prod_{j=1}^{l(\sigma_{d})}u_{d_{j}})z^{d}$ 
with length $l(\sigma_{d})\geq m+1$
and this proves (\ref{comb3}). Q.E.D\\

From Lemma 1, we can explicitly construct the mirror map,
\begin{eqnarray}
x(t)&=&t+\sum_{d=1}^{\infty}\frac{u_{d}}{d}e^{dt},\no\\
  u_{d}&=&Res_{z=0}(z^{-d-1}\exp(-d(\sum_{j=1}^{\infty}\frac{b_{j}}{j}z^{j}))),
\label{map1}
\end{eqnarray}
and we also have 
\begin{eqnarray}
b_{d}&=&Res_{z=0}(z^{-d-1}\exp(-d(\sum_{j=1}^{\infty}\frac{u_{j}}{j}z^{j}))).
\label{map2}
\end{eqnarray}
\begin{lem}
\begin{equation}
e^{dx(t)}=e^{dt}(1+d(\sum_{f=1}^{\infty}e^{ft}\sum_{\sigma_{f}\in P_{f}}
(-1)^{l(\sigma_{f})}(d+f)^{l(\sigma_{f})-1}S(\sigma_{f})
\prod_{j=1}^{l(\sigma_{f})}\frac{b_{d_{j}}}{d_{j}})).
\label{key}
\end{equation}
\end{lem}
{\it proof)}\\
We prove more generally that the identity
\begin{equation}
e^{\alpha x(t)}=e^{\alpha t}
(1+\alpha(\sum_{f=1}^{\infty}e^{ft}\sum_{\sigma_{f}\in P_{f}}
(-1)^{l(\sigma_{f})}(\alpha+f)^{l(\sigma_{f})-1}S(\sigma_{f})
\prod_{j=1}^{l(\sigma_{f})}\frac{b_{d_{j}}}{d_{j}}))
\label{genk}
\end{equation}
holds for arbitrary $\alpha\in C$. At first, we rewrite (\ref{genk}) into 
the form,  
\begin{equation}
\frac{1}{\alpha}(\exp(\alpha(x(t)-t))-1)
=\sum_{f=1}^{\infty}e^{ft}\sum_{\sigma_{f}\in P_{f}}
(-1)^{l(\sigma_{f})}(\alpha+f)^{l(\sigma_{f})-1}S(\sigma_{f})
\prod_{j=1}^{l(\sigma_{f})}\frac{b_{d_{j}}}{d_{j}}.
\label{ruse}
\end{equation}
Then using the fact that $t=t(x)$ and $x(t(x))=x$, we deduce 
from (\ref{ruse}),
\begin{eqnarray}
&&\frac{1}{\alpha}(\exp(\alpha(x-t(x)))-1)
=\sum_{f=1}^{\infty}e^{ft(x)}\sum_{\sigma_{f}\in P_{f}}
(-1)^{l(\sigma_{f})}(\alpha+f)^{l(\sigma_{f})-1}S(\sigma_{f})
\prod_{j=1}^{l(\sigma_{f})}\frac{b_{d_{j}}}{d_{j}}\no\\
&&\Longleftrightarrow \no\\
&&\frac{1}{\alpha}(\exp(-\alpha(\sum_{j=1}^{\infty}\frac{b_{j}}{j}e^{jx}))-1)=
\sum_{f=1}^{\infty}e^{fx}\exp(f(\sum_{j=1}^{\infty}\frac{b_{j}}{j}e^{jx}))
\sum_{\sigma_{f}\in P_{f}}
(-1)^{l(\sigma_{f})}(\alpha+f)^{l(\sigma_{f})-1}S(\sigma_{f})
\prod_{j=1}^{l(\sigma_{f})}\frac{b_{d_{j}}}{d_{j}}.\no\\
\label{nice}
\end{eqnarray}
By expanding $\exp(*)$ in (\ref{nice}), we obtain the following 
combinatorial relation among $S(\sigma_{m})$ 
that is equivalent to the statement of 
the lemma:  
\begin{eqnarray}
&&(-1)^{l(\sigma_{d})}\alpha^{l(\sigma_{d})-1}S(\sigma_{d})
=\sum_{f=0}^{d}
\sum_{{\sigma_{f}\in P_{f},\sigma_{d-f}\in P_{d-f}}\atop{\sigma_{f}\cup
\sigma_{d-f}=\sigma_{d}}}(-1)^{l(\sigma_{f})}f^{l(\sigma_{d-f})}
(\alpha+f)^{l(\sigma_{f})-1}
S(\sigma_{f})S(\sigma_{d-f})\no\\
&&=\sum_{j=0}^{l(\sigma_{d})-1}\alpha^{j}
\sum_{f=0}^{d}
\sum_{{\sigma_{f}\in P_{f},\sigma_{d-f}\in P_{d-f}}\atop{\sigma_{f}\cup
\sigma_{d-f}=\sigma_{d}}}(-1)^{l(\sigma_{f})}f^{l(\sigma_{d})-1-j}
{}_{l(\sigma_{f})-1}C_{j}S(\sigma_{f})S(\sigma_{d-f}).
\label{impov}
\end{eqnarray}
So we have only to prove,
\begin{equation}
\sum_{f=0}^{d}
\sum_{{\sigma_{f}\in P_{f},\sigma_{d-f}\in P_{d-f}}\atop{\sigma_{f}\cup
\sigma_{d-f}=\sigma_{d}}}(-1)^{l(\sigma_{f})}
{}_{l(\sigma_{f})-1}C_{l(\sigma_{d})-1}S(\sigma_{f})S(\sigma_{d-f}) 
=(-1)^{l(\sigma_{d})}S(\sigma_{d})
\label{reduce1}
\end{equation}
and
\begin{eqnarray} 
&&\frac{1}{j!}\sum_{f=0}^{d}
\sum_{{\sigma_{f}\in P_{f},\sigma_{d-f}\in P_{d-f}}\atop{\sigma_{f}\cup
\sigma_{d-f}=\sigma_{d}}}(-1)^{l(\sigma_{f})}f^{l(\sigma_{d})-1-j}
(\prod_{k=1}^{j}(l(\sigma_{f})-k))S(\sigma_{f})S(\sigma_{d-f}) 
=0,\no\\
&&(0\leq j \leq l(\sigma_{d})-2).
\label{reduce2}
\end{eqnarray}
As we did in proving Lemma 1, we can unify (\ref{reduce1})and (\ref{reduce2}) 
into a general statement:
\begin{eqnarray} 
&&\sum_{f=0}^{d}
\sum_{{\sigma_{f}\in P_{f},\sigma_{d-f}\in P_{d-f}}\atop{\sigma_{f}\cup
\sigma_{d-f}=\sigma_{d}}}(-1)^{l(\sigma_{f})}f^{m}
(\prod_{k=1}^{j}(l(\sigma_{f})-k))S(\sigma_{f})S(\sigma_{d-f}) 
=0,\no\\
&&(0\leq m+j\leq l(\sigma_{d})-1).
\label{sub}
\end{eqnarray}
The l.h.s. of (\ref{sub}) is coefficient of 
$(\prod_{j=1}^{l(\sigma_{d})}s_{d_{j}})z^{d}$ of generating function,
\begin{equation}
((z\frac{d}{dz})^{m}(\frac{d}{d\epsilon})^{j}
(\frac{1}{\epsilon}\exp(-\epsilon(\sum_{j=1}^{\infty}s_{j}z^{j})))|
_{\epsilon=1})\exp(\sum_{j=1}^{\infty}s_{j}z^{j}),
\label{last}
\end{equation}
but this does not contain monomials with 
length $l(\sigma_{d})\geq m+j+1$.
So (\ref{sub}) holds. Q.E.D. 
 
Using Lemma 2, we will now prove Theorem 2.
From (\ref{frac}) and (\ref{key}), we can see that Theorem 2 is 
equivalent to the combinatorial relation
\begin{eqnarray}
c_{d-j}&+&\sum_{g=j}^{d-1}g\cdot c_{g-j}
(\sum_{\sigma_{d-g}\in P_{d-g}}(-1)^{l(\sigma_{d-g})}d^{l(\sigma_{d-g})-1}
S(\sigma_{d-g})\prod_{i=1}^{l(\sigma_{d-g})}\frac{b_{d_{i}}}{d_{i}})\no\\
&=&\sum_{\sigma_{d-j}\in P_{d-j}}(-1)^{l(\sigma_{d-j})}d^{l(\sigma_{d-j})}
S(\sigma_{d-j})\prod_{i=1}^{l(\sigma_{d-j})}\frac{b_{d_{i}}}{d_{i}}.
\label{rewr}
\end{eqnarray}
Let us denote as
\begin{equation}
\exp(-d(\sum_{j=1}^{\infty}\frac{b_{j}}{j}z^{j}))=
\sum_{m=0}^{\infty}\alpha_{m,-d}z^{m}.
\label{simple}
\end{equation}
Then we can rewrite (\ref{rewr}) as 
\begin{equation}
c_{d-j}+\sum_{g=j}^{d-1}(\frac{g}{d})\cdot c_{g-j}
\cdot\alpha_{d-g,-d}=\alpha_{d-j,-d}.
\label{rewr2}
\end{equation}
We apply $z\frac{d}{dz}$ to
the both sides of (\ref{simple}):
\begin{equation}
(-d(\sum_{i=1}^{\infty}b_{i}z^{i}))
\exp(-d(\sum_{j=1}^{\infty}\frac{b_{j}}{j}z^{j}))=
\sum_{m=0}^{\infty}m\cdot \alpha_{m,-d}z^{m}.
\label{msimple}
\end{equation}
So, we can easily derive 
\begin{equation}
\sum_{m=0}^{\infty}\alpha_{m,-d}\cdot z^{m}=
\sum_{m=0}^{\infty}c_{m}z^{m}+\sum_{m=1}^{\infty}
\sum_{l=1}^{m}(\frac{d-l}{d})\cdot\alpha_{l,-d}\cdot c_{m-l}
\cdot z^{m}
\label{form}
\end{equation}
Substituting $l=d-g$, we see that (\ref{form}) is nothing but 
(\ref{rewr2}). Q.E.D.
\newpage
\section*{ Appendix B:  Recursive Formulas for Fano Hypersurfaces
with $c_{1}(M_{N}^{k})\geq 2$}
\begin{eqnarray}
L^{N,k,1}_{m}&=&L^{N+1,k,1}_{m}:=L^{k}_{m}\\
L^{N,k,2}_{m}&=&\frac{1}{2}(L^{N+1,k,2}_{m-1}+L^{N+1,k,2}_{m}
+2L^{N+1,k,1}_{m}\cdot L^{N+1,k,1}_{m+(N-k)})\\
L^{N,k,3}_{m}&=&\frac{1}{18}(4L^{N+1,k,3}_{m-2}+10L^{N+1,k,3}_{m-1}
+4L^{N+1,k,3}_{m}\nonumber\\
&&+12L^{N+1,k,2}_{m-1}\cdot L^{N+1,k,1}_{m+2(N-k)}
+9L^{N+1,k,2}_{m}\cdot L^{N+1,k,1}_{m+2(N-k)}\nonumber\\
&&+6L^{N+1,k,2}_{m}\cdot L^{N+1,k,1}_{m+1+2(N-k)}\nonumber\\
&&+6 L^{N+1,k,1}_{m-1}\cdot L^{N+1,k,2}_{m-1+(N-k)}
+9 L^{N+1,k,1}_{m}\cdot L^{N+1,k,2}_{m-1+(N-k)}\nonumber\\
&&+12 L^{N+1,k,1}_{m}\cdot L^{N+1,k,2}_{m+(N-k)}\nonumber\\
&&+18L^{N+1,k,1}_{m}\cdot L^{N+1,k,1}_{m+(N-k)}\cdot L^{N+1,k,1}_{m+2(N-k)})
\label{rec0}
\end{eqnarray}
In the following, we omit $N+1,k$ of $L_{n}^{N+1,k,4}$ in the r.h.s. 
for brevity.
\begin{eqnarray}
L_{n}^{N,k,4}
&=&\frac{1}{32}(3L_{n-3}^{4}+13L_{n-2}^{4}
+13L_{n-1}^{4}+3L_{n}^{4})\nonumber\\
&&+\frac{1}{72}(9L_{n-2}^{1}L_{n-2+N-k}^{3}+12L_{n-1}^{1}L_{n-2+N-k}^{3}
+16L_{n}^{1}L_{n-2+N-k}^{3}\nonumber\\
&&+36L_{n-1}^{1}L_{n-1+N-k}^{3}
+44L_{n}^{1}L_{n-1+N-k}^{3}+27L_{n}^{1}L_{n+N-k}^{3})\nonumber\\
&&+\frac{1}{16}(3L_{n-2}^{2}L_{n-1+2(N-k)}^{2}
+6L_{n-1}^{2}L_{n-1+2(N-k)}^{2}
+4L_{n}^{2}L_{n-1+2(N-k)}^{2}\nonumber\\
&&+10L_{n-1}^{2}L_{n+2(N-k)}^{2}
+6L_{n}^{2}L_{n+2(N-k)}^{2}+3L_{n}^{2}L_{n+1+2(N-k)}^{2})
\nonumber\\
&&+\frac{1}{72}(27L_{n-2}^{3}L_{n+3(N-k)}^{1}
+44L_{n-1}^{3}L_{n+3(N-k)}^{1}
+16L_{n}^{3}L_{n+3(N-k)}^{1}\nonumber\\
&&+36L_{n-1}^{3}L_{n+1+3(N-k)}^{1}
+12L_{n}^{3}L_{n+1+3(N-k)}^{1}+
9L_{n}^{3}L_{n+2+3(N-k)}^{1})\nonumber\\
&&+\frac{1}{12}(3L_{n-1}^{1}L_{n-1+N-k}^{1}L_{n-1+2(N-k)}^{2}
+4L_{n}^{1}L_{n-1+N-k}^{1}L_{n-1+2(N-k)}^{2}\nonumber\\
&&+6L_{n}^{1}L_{n+N-k}^{1}L_{n-1+2(N-k)}^{2}
+9L_{n}^{1}L_{n+N-k}^{1}L_{n+2(N-k)}^{2})\nonumber\\
&&+\frac{1}{6}(3L_{n-1}^{1}L_{n-1+N-k}^{2}L_{n+3(N-k)}^{1}
+4L_{n}^{1}L_{n-1+N-k}^{2}L_{n+3(N-k)}^{1}\nonumber\\
&&+4L_{n}^{1}L_{n+N-k}^{2}L_{n+3(N-k)}^{1}
+3L_{n}^{1}L_{n+N-k}^{2}L_{n+1+3(N-k)}^{1})\nonumber\\
&&+\frac{1}{12}(9L_{n-1}^{2}L_{n+2(N-k)}^{1}L_{n+3(N-k)}^{1}
+6L_{n}^{2}L_{n+2(N-k)}^{1}L_{n+3(N-k)}^{1}\nonumber\\
&&+4L_{n}^{2}L_{n+1+2(N-k)}^{1}L_{n+3(N-k)}^{1}
+3L_{n}^{2}L_{n+1+2(N-k)}^{1}L_{n+1+3(N-k)}^{1})\nonumber\\
&&+L_{n}^{1}L_{n+N-k}^{1}L_{n+2(N-k)}^{1}L_{n+3(N-k)}^{1}.    
\end{eqnarray}
\newpage
\begin{eqnarray}
&&L^{N,k,5}_{n} \nonumber\\
&=&\frac{24}{625}L_{n-4}^{5}+\frac{154}{625}L_{n-3}^{5}
+\frac{269}{625}L_{n-2}^{5}+\frac{154}{625}L_{n-1}^{5}
+\frac{24}{625}L_{n}^{5} \nonumber\\
&&+\frac{6}{125}L_{n-3}^{1}L_{n-3+N-k}^{4}
+\frac{3}{50}L_{n-2}^{1}L_{n-3+N-k}^{4}
+\frac{3}{40}L_{n-1}^{1}L_{n-3+N-k}^{4}\nonumber\\
&&+\frac{3}{32}L_{n}^{1}L_{n-3+N-k}^{4}
+\frac{37}{125}L_{n-2}^{1}L_{n-2+N-k}^{4}
+\frac{71}{200}L_{n-1}^{1}L_{n-2+N-k}^{4}\nonumber\\
&&+\frac{17}{40}L_{n}^{1}L_{n-2+N-k}^{4}
+\frac{58}{125}L_{n-1}^{1}L_{n-1+N-k}^{4}
+\frac{393}{800}L_{n}^{1}L_{n-1+N-k}^{4}\nonumber\\
&&+\frac{24}{125}L_{n}^{1}L_{n+N-k}^{4}\nonumber\\
&&+\frac{8}{125}L_{n-3}^{2}L_{n-2+2(N-k)}^{3}+
\frac{8}{75}L_{n-2}^{2}L_{n-2+2(N-k)}^{3}
+\frac{8}{45}L_{n-1}^{2}L_{n-2+2(N-k)}^{3}\nonumber\\
&&+\frac{1}{9}L_{n}^{2}L_{n-2+2(N-k)}^{3} 
+\frac{46}{125}L_{n-2}^{2}L_{n-1+2(N-k)}^{3}
+\frac{122}{225}L_{n-1}^{2}L_{n-1+2(N-k)}^{3}\nonumber\\
&&+\frac{29}{90}L_{n}^{2}L_{n-1+2(N-k)}^{3}+
\frac{59}{125}L_{n-1}^{2}L_{n+2(N-k)}^{3}
+\frac{6}{25}L_{n}^{2}L_{n+2(N-k)}^{3}\nonumber\\
&&+\frac{12}{125}L_{n}^{2}L_{n+1+2(N-k)}^{3}\nonumber\\
&&+\frac{12}{125}L_{n-3}^{3}L_{n-1+3(N-k)}^{2}+
\frac{6}{25}L_{n-2}^{3}L_{n-1+3(N-k)}^{2}
+\frac{29}{90}L_{n-1}^{3}L_{n-1+3(N-k)}^{2}\nonumber\\
&&+\frac{1}{9}L_{n}^{3}L_{n-1+3(N-k)}^{2}
+\frac{59}{125}L_{n-2}^{3}L_{n+3(N-k)}^{2}
+\frac{122}{225}L_{n-1}^{3}L_{n+3(N-k)}^{2}\nonumber\\
&&+\frac{8}{45}L_{n}^{3}L_{n+3(N-k)}^{2}
+\frac{46}{125}L_{n-1}^{3}L_{n+1+3(N-k)}^{2}
+\frac{8}{75}L_{n}^{3}L_{n+1+3(N-k)}^{2}\nonumber\\
&&+\frac{8}{125}L_{n}^{3}L_{n+2+3(N-k)}^{2}\nonumber\\
&&+\frac{24}{125}L_{n-3}^{4}L_{n+4(N-k)}^{1}
+\frac{393}{800}L_{n-2}^{4}L_{n+4(N-k)}^{1}
+\frac{17}{40}L_{n-1}^{4}L_{n+4(N-k)}^{1}\nonumber\\
&&+\frac{3}{32}L_{n}^{4}L_{n+4(N-k)}^{1} 
+\frac{58}{125}L_{n-2}^{4}L_{n+1+4(N-k)}^{1}
+\frac{71}{200}L_{n-1}^{4}L_{n+1+4(N-k)}^{1}\nonumber\\
&&+\frac{3}{40}L_{n}^{4}L_{n+1+4(N-k)}^{1}
+\frac{37}{125}L_{n-1}^{4}L_{n+2+4(N-k)}^{1}
+\frac{3}{50}L_{n}^{4}L_{n+2+4(N-k)}^{1}\nonumber\\
&&+\frac{6}{125}L_{n}^{4}L_{n+3+4(N-k)}^{1}\nonumber\\
&&+\frac{2}{25}L_{n-2}^{1}L_{n-2+N-k}^{1}L_{n-2+2(N-k)}^{3}
+\frac{1}{10}L_{n-1}^{1}L_{n-2+N-k}^{1}L_{n-2+2(N-k)}^{3}\nonumber\\
&&+\frac{1}{8}L_{n}^{1}L_{n-2+N-k}^{1}L_{n-2+2(N-k)}^{3}
+\frac{2}{15}L_{n-1}^{1}L_{n-1+N-k}^{1}L_{n-2+2(N-k)}^{3}\nonumber\\
&&+\frac{1}{6}L_{n}^{1}L_{n-1+N-k}^{1}L_{n-2+2(N-k)}^{3}
+\frac{2}{9}L_{n}^{1}L_{n+N-k}^{1}L_{n-2+2(N-k)}^{3}\nonumber\\
&&+\frac{11}{25}L_{n-1}^{1}L_{n-1+N-k}^{1}L_{n-1+2(N-k)}^{3}
+\frac{21}{40}L_{n}^{1}L_{n-1+N-k}^{1}L_{n-1+2(N-k)}^{3}\nonumber\\
&&+\frac{29}{45}L_{n}^{1}L_{n+N-k}^{1}L_{n-1+2(N-k)}^{3}
+\frac{12}{25}L_{n}^{1}L_{n+N-k}^{1}L_{n+2(N-k)}^{3}\nonumber\\
&&+\frac{6}{25}L_{n-2}^{1}L_{n-2+N-k}^{3}L_{n+4(N-k)}^{1}
+\frac{3}{10}L_{n-1}^{1}L_{n-2+N-k}^{3}L_{n+4(N-k)}^{1}\nonumber\\
&&+\frac{3}{8}L_{n}^{1}L_{n-2+N-k}^{3}L_{n+4(N-k)}^{1}
+\frac{23}{40}L_{n-1}^{1}L_{n-1+N-k}^{3}L_{n+4(N-k)}^{1}\nonumber\\
&&+\frac{2}{3}L_{n}^{1}L_{n-1+N-k}^{3}L_{n+4(N-k)}^{1}
+\frac{3}{8}L_{n}^{1}L_{n+N-k}^{3}L_{n+4(N-k)}^{1}\nonumber\\
&&+\frac{13}{25}L_{n-1}^{1}L_{n-1+N-k}^{3}L_{n+1+4(N-k)}^{1}
+\frac{23}{40}L_{n}^{1}L_{n-1+N-k}^{3}L_{n+1+4(N-k)}^{1}\nonumber\\
&&+\frac{3}{10}L_{n}^{1}L_{n+N-k}^{3}L_{n+1+4(N-k)}^{1}
+\frac{6}{25}L_{n}^{1}L_{n+N-k}^{3}L_{n+2+4(N-k)}^{1}\nonumber\\
&&+\frac{12}{25}L_{n-2}^{3}L_{n+3(N-k)}^{1}L_{n+4(N-k)}^{1}
+\frac{29}{45}L_{n-1}^{3}L_{n+3(N-k)}^{1}L_{n+4(N-k)}^{1}\nonumber\\
&&+\frac{2}{9}L_{n}^{3}L_{n+3(N-k)}^{1}L_{n+4(N-k)}^{1}
+\frac{21}{40}L_{n-1}^{3}L_{n+1+3(N-k)}^{1}L_{n+4(N-k)}^{1}\nonumber\\
&&+\frac{1}{6}L_{n}^{3}L_{n+1+3(N-k)}^{1}L_{n+4(N-k)}^{1}
+\frac{1}{8}L_{n}^{3}L_{n+2+3(N-k)}^{1}L_{n+4(N-k)}^{1}\nonumber\\
&&+\frac{11}{25}L_{n-1}^{3}L_{n+1+3(N-k)}^{1}L_{n+1+4(N-k)}^{1}
+\frac{2}{15}L_{n}^{3}L_{n+1+3(N-k)}^{1}L_{n+1+4(N-k)}^{1}\nonumber\\
&&+\frac{1}{10}L_{n}^{3}L_{n+2+3(N-k)}^{1}L_{n+1+4(N-k)}^{1}
+\frac{2}{25}L_{n}^{3}L_{n+2+3(N-k)}^{1}L_{n+2+4(N-k)}^{1}\nonumber\\
&&+\frac{3}{25}L_{n-2}^{1}L_{n-2+N-k}^{2}L_{n-1+3(N-k)}^{2}
+\frac{3}{20}L_{n-1}^{1}L_{n-2+N-k}^{2}L_{n-1+3(N-k)}^{2}\nonumber\\
&&+\frac{3}{16}L_{n}^{1}L_{n-2+N-k}^{2}L_{n-1+3(N-k)}^{2}
+\frac{3}{10}L_{n-1}^{1}L_{n-1+N-k}^{2}L_{n-1+3(N-k)}^{2}\nonumber\\
&&+\frac{3}{8}L_{n}^{1}L_{n-1+N-k}^{2}L_{n-1+3(N-k)}^{2}
+\frac{1}{3}L_{n}^{1}L_{n+N-k}^{2}L_{n-1+3(N-k)}^{2}\nonumber\\
&&+\frac{14}{25}L_{n-1}^{1}L_{n-1+N-k}^{2}L_{n+3(N-k)}^{2}
+\frac{53}{80}L_{n}^{1}L_{n-1+N-k}^{2}L_{n+3(N-k)}^{2}\nonumber\\
&&+\frac{8}{15}L_{n}^{1}L_{n+N-k}^{2}L_{n+3(N-k)}^{2}
+\frac{8}{25}L_{n}^{1}L_{n+N-k}^{2}L_{n+1+3(N-k)}^{2}\nonumber\\
&&+\frac{4}{25}L_{n-2}^{2}L_{n-1+2(N-k)}^{1}L_{n-1+3(N-k)}^{2}
+\frac{4}{15}L_{n-1}^{2}L_{n-1+2(N-k)}^{1}L_{n-1+3(N-k)}^{2}\nonumber\\
&&+\frac{1}{6}L_{n}^{2}L_{n-1+2(N-k)}^{1}L_{n-1+3(N-k)}^{2}
+\frac{2}{5}L_{n-1}^{2}L_{n+2(N-k)}^{1}L_{n-1+3(N-k)}^{2}\nonumber\\
&&+\frac{1}{4}L_{n}^{2}L_{n+2(N-k)}^{1}L_{n-1+3(N-k)}^{2}
+\frac{1}{6}L_{n}^{2}L_{n+1+2(N-k)}^{1}L_{n-1+3(N-k)}^{2}\nonumber\\
&&+\frac{17}{25}L_{n-1}^{2}L_{n+2(N-k)}^{1}L_{n+3(N-k)}^{2}
+\frac{2}{5}L_{n}^{2}L_{n+2(N-k)}^{1}L_{n+3(N-k)}^{2}\nonumber\\
&&+\frac{4}{15}L_{n}^{2}L_{n+1+2(N-k)}^{1}L_{n+3(N-k)}^{2}
+\frac{4}{25}L_{n}^{2}L_{n+1+2(N-k)}^{1}L_{n+1+3(N-k)}^{2}\nonumber\\
&&+\frac{8}{25}L_{n-2}^{2}L_{n-1+2(N-k)}^{2}L_{n+4(N-k)}^{1}
+\frac{8}{15}L_{n-1}^{2}L_{n-1+2(N-k)}^{2}L_{n+4(N-k)}^{1}\nonumber\\
&&+\frac{1}{3}L_{n}^{2}L_{n-1+2(N-k)}^{2}L_{n+4(N-k)}^{1}
+\frac{53}{80}L_{n-1}^{2}L_{n+2(N-k)}^{2}L_{n+4(N-k)}^{1}\nonumber\\
&&+\frac{3}{8}L_{n}^{2}L_{n+2(N-k)}^{2}L_{n+4(N-k)}^{1}
+\frac{3}{16}L_{n}^{2}L_{n+1+2(N-k)}^{2}L_{n+4(N-k)}^{1}\nonumber\\
&&+\frac{14}{25}L_{n-1}^{2}L_{n+2(N-k)}^{2}L_{n+1+4(N-k)}^{1}
+\frac{3}{10}L_{n}^{2}L_{n+2(N-k)}^{2}L_{n+1+4(N-k)}^{1}\nonumber\\
&&+\frac{3}{20}L_{n}^{2}L_{n+1+2(N-k)}^{2}L_{n+1+4(N-k)}^{1}
+\frac{3}{25}L_{n}^{2}L_{n+1+2(N-k)}^{2}L_{n+2+4(N-k)}^{1}\nonumber\\
&&+\frac{1}{5}L_{n-1}^{1}L_{n-1+N-k}^{1}L_{n-1+2(N-k)}^{1}L_{n-1+3(N-k)}^{2}
+\frac{1}{4}L_{n}^{1}L_{n-1+N-k}^{1}L_{n-1+2(N-k)}^{1}
L_{n-1+3(N-k)}^{2}\nonumber\\
&&+\frac{1}{3}L_{n}^{1}L_{n+N-k}^{1}L_{n-1+2(N-k)}^{1}L_{n-1+3(N-k)}^{2}
+\frac{1}{2} L_{n}^{1}L_{n+N-k}^{1}L_{n+2(N-k)}^{1}
L_{n-1+3(N-k)}^{2}\nonumber\\
&&+\frac{4}{5}L_{n}^{1}L_{n+N-k}^{1}L_{n+2(N-k)}^{1}L_{n+3(N-k)}^{2}
+\frac{2}{5}  L_{n-1}^{1}L_{n-1+N-k}^{1}L_{n-1+2(N-k)}^{2}
L_{n+4(N-k)}^{1}\nonumber\\
&&+\frac{1}{2}L_{n}^{1}L_{n-1+N-k}^{1}L_{n-1+2(N-k)}^{2}L_{n+4(N-k)}^{1}
+\frac{2}{3} L_{n}^{1}L_{n+N-k}^{1}L_{n-1+2(N-k)}^{2}
L_{n+4(N-k)}^{1}\nonumber\\
&&+\frac{3}{4} L_{n}^{1}L_{n+N-k}^{1}L_{n+2(N-k)}^{2}L_{n+4(N-k)}^{1}
+\frac{3}{5}  L_{n}^{1}L_{n+N-k}^{1}L_{n+2(N-k)}^{2}
L_{n+1+4(N-k)}^{1}\nonumber\\
&&+\frac{3}{5}L_{n-1}^{1}L_{n-1+N-k}^{2}L_{n+3(N-k)}^{1}L_{n+4(N-k)}^{1}
+\frac{3}{4}L_{n}^{1}L_{n-1+N-k}^{2}L_{n+3(N-k)}^{1}L_{n+4(N-k)}^{1}\nonumber\\
&&+\frac{2}{3}L_{n}^{1}L_{n+N-k}^{2}L_{n+3(N-k)}^{1}L_{n+4(N-k)}^{1}
+\frac{1}{2}L_{n}^{1}L_{n+N-k}^{2}L_{n+1+3(N-k)}^{1}L_{n+4(N-k)}^{1}\nonumber\\
&&+\frac{2}{5}L_{n}^{1}L_{n+N-k}^{2}L_{n+1+3(N-k)}^{1}L_{n+1+4(N-k)}^{1}
+\frac{4}{5}L_{n-1}^{2}L_{n+2(N-k)}^{1}L_{n+3(N-k)}^{1}
L_{n+4(N-k)}^{1}\nonumber\\
&&+\frac{1}{2}L_{n}^{2}L_{n+2(N-k)}^{1}L_{n+3(N-k)}^{1}L_{n+4(N-k)}^{1}
+\frac{1}{3}L_{n}^{2}L_{n+1+2(N-k)}^{1}L_{n+3(N-k)}^{1}
L_{n+4(N-k)}^{1}\nonumber\\
&&+\frac{1}{4}L_{n}^{2}L_{n+1+2(N-k)}^{1}L_{n+1+3(N-k)}^{1}L_{n+4(N-k)}^{1}
+\frac{1}{5}L_{n}^{2}L_{n+1+2(N-k)}^{1}L_{n+1+3(N-k)}^{1}
L_{n+1+4(N-k)}^{1}\nonumber\\
&&+L_{n}^{1}L_{n+N-k}^{1}L_{n+2(N-k)}^{1}
L_{n+3(N-k)}^{1}L_{n+4(N-k)}^{1}
\label{quin}    
\end{eqnarray}
\newpage
\begin{table}[p]
\caption{Initial Condition of Recursive Formulas
(Beauville's Region)}
\begin{center}
\begin{tabular}{|l|l l l|}
\hline
 $k=5$ & $L_{0}^{10,5,1}=120$ & $L_{1}^{10,5,1}=770 $ &
$L_{2}^{16,8,1}=1345 $\\
& $ L_{3}^{10,5,1}=770 $ & $ L_{4}^{10,5,1}=120 $ &\\ 
\hline
 $k=6$ & $L_{0}^{12,6,1}=720$ & $L_{1}^{12,6,1}=6264 $ &
$L_{2}^{12,6,1}=16344 $\\
& $ L_{3}^{12,6,1}=16344$ & $ L_{4}^{12,6,1}=6264 $ &
$ L_{5}^{12,6,1}=720 $\\
\hline
 $k=7$ &$L_{0}^{14,7,1}=5040$ & $L_{1}^{14,7,1}=56196 $ &
$L_{2}^{14,7,1}=200452  $\\
& $ L_{3}^{14,7,1}=300167 $ & $ L_{4}^{14,7,1}=200452 $ &
$ L_{5}^{14,7,1}=56196 $\\
&$L_{6}^{14,7,1}=5040 $  & & \\ 
\hline
 $k=8$  & $L_{0}^{16,8,1}=40320$ & $L_{1}^{16,8,1}=554112 $ &
$L_{2}^{16,8,1}=2552192 $\\
& $ L_{3}^{16,8,1}=5241984 $ & $ L_{4}^{16,8,1}=5241984 $ &
$ L_{5}^{16,8,1}=2552192 $\\
&$L_{6}^{16,8,1}=554112$  & $ L_{7}^{16,8,1}=40320$ & \\
\hline
$ k=9 $&$L_{0}^{18,9,1}=362880 $ & $  L_{1}^{18,9,1}=5973264 $ & 
$ L_{2}^{18,9,1}=34138908 $\\
& $ L_{3}^{18,9,1}=90857052 $ & $ L_{4}^{18,9,1}=124756281 $ &
$ L_{5}^{18,9,1}=90857052 $\\
&$ L_{6}^{18,9,1}=34138908 $ & $ L_{7}^{18,9,1}=5973264 $ &
$ L_{8}^{18,9,1}=362880 $\\
\hline
$ k=10 $ &$L_{0}^{20,10,1}=3628800$ &$L_{1}^{20,10,1}=69998400$ &
$L_{2}^{20,10,1}=482076000$\\
 &$L_{3}^{20,10,1}=1597493600$ &$L_{4}^{20,10,1}=2846803200 $ &
$L_{5}^{20,10,1}=2846803200 $\\
 &$L_{6}^{20,10,1}=1597493600$ &$L_{7}^{20,10,1}=482076000$ &
$L_{8}^{20,10,1}=69998400 $\\
 &$L_{9}^{20,10,1}=3628800$ & & \\ 
\hline
$ k=11 $ &$L_{0}^{22,11,1}=39916800 $ &$L_{1}^{22,11,1}=886897440$ &
$L_{2}^{22,11,1}=7196676696$\\ 
&$L_{3}^{22,11,1}=28831752092 $ &$L_{4}^{22,11,1}=64088868338 $ &
$L_{5}^{22,11,1}=83223447879 $\\
&$L_{6}^{22,11,1}=64088868338 $ &$L_{7}^{22,11,1}=28831752092 $ &
$L_{8}^{22,11,1}=7196676696 $\\
&$L_{9}^{22,11,1}=886897440 $ &$L_{10}^{22,11,1}=39916800 $ & \\
\hline
$ k=12 $ &$L_{0}^{24,12,1}=479001600 $ &$L_{1}^{24,12,1}=12089295360 $ &
$L_{2}^{24,12,1}=113548220928 $\\
&$L_{3}^{24,12,1}=537643920384 $ &$L_{4}^{24,12,1}=1447536199680 $ &
$L_{5}^{24,12,1}=2346753586176 $\\
&$L_{6}^{24,12,1}=2346753586176 $ &$L_{7}^{24,12,1}=1447536199680 $ &
$L_{8}^{24,12,1}=537643920384 $\\
&$L_{9}^{24,12,1}=113548220928 $ &$L_{10}^{24,12,1}=12089295360 $ &
$L_{11}^{24,12,1}=479001600 $\\
\hline
$ k=13 $ &$L_{0}^{26,13,1}=6227020800 $ 
&$L_{1}^{26,13,1}=176484597120 $ &
$L_{2}^{26,13,1}=1891322394624 $\\
&$L_{3}^{26,13,1}=10395857852328 $ 
&$L_{4}^{26,13,1}=33141735509116 $ &
$L_{5}^{26,13,1}=65146083016534 $\\
&$L_{6}^{26,13,1}=81359685811209 $ 
&$L_{7}^{26,13,1}=65146083016534 $ &
$L_{8}^{26,13,1}=33141735509116 $\\
&$L_{9}^{26,13,1}=10395857852328 $ 
&$L_{10}^{26,13,1}=1891322394624 $ &
$L_{11}^{26,13,1}=176484597120 $\\
&$L_{12}^{26,13,1}=6227020800$  & &\\
\hline
$k=14 $ &$L_{0}^{28,14,1}=87178291200$ 
&$L_{1}^{28,14,1}=2748022986240 $ &
$L_{2}^{28,14,1}=33205209053184 $\\
&$L_{3}^{28,14,1}=208813330975872 $ 
&$L_{4}^{28,14,1}=774174722002304 $ &
$L_{5}^{28,14,1}=1803924774789760 $\\
&$L_{6}^{28,14,1}=2733050174680448  $ 
&$L_{7}^{28,14,1}= 2733050174680448  $ &
$L_{8}^{28,14,1}=1803924774789760 $\\
&$L_{9}^{28,14,1}=774174722002304 $ 
&$L_{10}^{28,14,1}=208813330975872 $ &
$L_{11}^{28,14,1}=33205209053184 $\\
&$L_{12}^{28,14,1}=2748022986240  $ 
&$L_{13}^{28,14,1}=87178291200 $ &\\
\hline
$ k=15 $&$L_{0}^{30,15,1}=1307674368000$ 
&$L_{1}^{30,15,1}=45472329504000  $ &
$L_{2}^{30,15,1}=613390541616000 $\\
&$L_{3}^{30,15,1}=4360309637094000 $ 
&$L_{4}^{30,15,1}=18527344278048000 $ &
$L_{5}^{30,15,1}=50264090344359000 $\\
&$L_{6}^{30,15,1}=90331361620677000 $ 
&$L_{7}^{30,15,1}=109607337529527375  $ &
$L_{8}^{30,15,1}=90331361620677000 $\\
&$L_{9}^{30,15,1}=50264090344359000 $ 
&$L_{10}^{30,15,1}=18527344278048000 $ &
$L_{11}^{30,15,1}=4360309637094000 $\\
&$L_{12}^{30,15,1}=613390541616000 $ 
&$L_{13}^{30,15,1}=45472329504000 $ &
$L_{14}^{30,15,1}=1307674368000 $\\
\hline
\multicolumn{4}{l}{(All the $L_n^{2k,k,d}\;(d\geq 2)$'s are zero.)}\\
\end{tabular}
\end{center}
\end{table}
\newpage
\begin{table}
\caption{Examples of the Generalized Mirror Transformation for 
the $d=3,4$ Cases}
\begin{center}
\begin{tabular}{l}
%$L_{5}^{9,10,3}=344441697426824242113176000000/3$\\
%$=2458271442602551980414496000000/9-978845914797711314643808000000/9-$\\
%$3*482076000*(28650399898527120000-5669679998699100000)+$\\
%$(9/2)*482076000*482076000*(2846803200+2846803200-1597493600-482076000)-$\\
%$(3/2)*5669679998699100000*(2846803200-482076000)$\\
%$L_{6}^{9,10,4}=19426605589905742065890532579007639843750/27$\\
%$=29756263013229634742238766406795989843750/9-$\\
%$6140535133309731446172327495292792187500/3-$\\
%$4*482076000*(2458271442602551980414496000000/9-$\\
%$978845914797711314643808000000/9)-$\\
%$2*5669679998699100000*(19011543456308490000-$\\
%$5669679998699100000)-$\\
%$(4/3)*978845914797711314643808000000/9*(1597493600-482076000)+$\\
%$8*482076000*482076000*(19011543456308490000+28650399898527120000-$\\
%$19011543456308490000-5669679998699100000)+$\\
%$8*482076000*5669679998699100000*(1597493600+2846803200-$\\
%$1597493600-482076000)-$\\
%$32/3*482076000*482076000*482076000*(2846803200+2846803200$\\
%$-1597493600-482076000)$\\
$L_{5}^{10,11,3}=7601114016412355265506038149776315/3$\\
$=39321013161929775850199104322696996/9-$\\
$11051810355155318663381224366483160/9-$\\
$3*7196676696*(40216130393485580917383/2-2153815059753018506790)+$\\
$(9/2)*7196676696*7196676696*(83223447879+64088868338-$\\
$28831752092-7196676696)-$\\
$(3/2)*2153815059753018506790*(83223447879-7196676696)$\\
\\
$L_{6}^{10,11,3}=13783958731158999754651957610916334/3$\\
$=61012348943229750362670547134032423/9-$\\
$11051810355155318663381224366483160/9-$\\
$3*7196676696*(40216130393485580917383/2+40216130393485580917383/2-$\\
$9451710952055403714441-2153815059753018506790+$\\
$64088868338*64088868338-64088868338*28831752092-83223447879*28831752092+$\\
$83223447879*7196676696-64088868338*7196676696+28831752092*28831752092)+$\\
$(9/2)*7196676696*7196676696*(64088868338+2*83223447879-$\\
$2*28831752092-7196676696)-$\\
$(3/2)*2153815059753018506790*(64088868338+83223447879-$\\
$28831752092-7196676696)$\\
\\
$L_{6}^{10,11,4}=173094363459727796215838406309670890661018005223/216$\\
$=538190721372744329415746872927006601586177865657/288-$\\
$73190750568118350991831897085289272131127171095/96-$\\
$4*7196676696*(61012348943229750362670547134032423/9-$\\
$11051810355155318663381224366483160/9)-$\\
$2*2153815059753018506790*(40216130393485580917383/2-$\\
$2153815059753018506790)-$\\
$(4/3)*11051810355155318663381224366483160/9*(64088868338-7196676696)+$\\
$8*7196676696*7196676696*(40216130393485580917383/2+$\\
$40216130393485580917383/2-$\\
$9451710952055403714441-2153815059753018506790+$\\
$64088868338*64088868338-64088868338*28831752092-83223447879*28831752092+$\\
$83223447879*7196676696-64088868338*7196676696+28831752092*28831752092)+$\\
$8*7196676696*2153815059753018506790*(64088868338+83223447879-$\\
$28831752092-7196676696)-$\\
$32/3*7196676696*7196676696*7196676696*(64088868338+2*83223447879-$\\
$2*28831752092-7196676696)$\\
\end{tabular}
\end{center}
\end{table}
\begin{table}
\caption{More Example of the Generalized Mirror Transformation for 
the  $d=3$ Case}
\begin{center}
\begin{tabular}{l}
$L_{9}^{13,15,3}=14356269698724856586166649497084995544762232083984375$\\
$=23334416034364889092865122924300384023499300787109375-$\\
$7349834530017719907441240703342049643620341898437500-$\\
$3*4360309637094000*(148632392734375058940601059038015625/4+$\\
$97609182646575948239192133869015625/2-16577238305553301159536028880531250-$\\
$4454602732448692602148965129468750+50264090344359000*50264090344359000-$\\
$18527344278048000*(50264090344359000+90331361620677000+109607337529527375+$\\
$90331361620677000+50264090344359000)+$\\
$4360309637094000*(90331361620677000+109607337529527375+90331361620677000)+$\\
$18527344278048000*(109607337529527375+90331361620677000+50264090344359000+$\\
$18527344278048000+4360309637094000)-$\\
$4360309637094000*(109607337529527375+90331361620677000+50264090344359000+$\\
$18527344278048000)+$\\
$50264090344359000*90331361620677000+90331361620677000*50264090344359000-$\\
$50264090344359000*(50264090344359000+90331361620677000+$\\
$109607337529527375+90331361620677000+50264090344359000)+$\\
$4360309637094000*109607337529527375-$\\
$109607337529527375*18527344278048000-90331361620677000*4360309637094000+$\\
$50264090344359000*(109607337529527375+90331361620677000+$\\
$50264090344359000+18527344278048000+4360309637094000)-$\\
$4360309637094000*50264090344359000)+$\\
$(9/2)*4360309637094000*4360309637094000*$\\
$(50264090344359000+2*90331361620677000+3*109607337529527375+$\\
$2*90331361620677000+50264090344359000-109607337529527375-$\\
$2*90331361620677000-3*50264090344359000-2*18527344278048000-$\\
$4360309637094000)-$\\
$(3/2)*4454602732448692602148965129468750*(50264090344359000+$\\
$90331361620677000-18527344278048000-4360309637094000)$\\
\end{tabular}
\end{center}
\end{table}

\end{document}